%% file: info41-5.tex
\documentclass[11pt,twoside]{article}
\usepackage{acta-info}
\usepackage{euler}

\newcommand{\vol}{\textbf{4,} 1 (2012)  67--118}   
\newcommand{\amss}{\amssubj{%
05C21
}}   
\newcommand{\CRs}{\CRsubj{%
G.2.2
}}   
\newcommand{\keyw}{\keywords{%
network, flow, algorithm, minimum-cost flow, efficient implementation
}}

\usepackage{array}
\usepackage{multirow}
\begin{document}

\title{Efficient implementations of minimum-cost flow algorithms}
\maketitle

\twoauthors{%
\href{http://www.cs.elte.hu/~kiraly}{Zolt\'an KIR\'ALY}%
}{%
\href{http://www.cs.elte.hu/index.html?lang=en}{Dept. of Computer Science} and\\%
\href{http://www.cs.elte.hu/egres/}{MTA-ELTE Egerv\'ary Research Group}\\%
\href{http://www.elte.hu/en}{E\"otv\"os Lor\'and University}\\%
Budapest, Hungary%
}{%
\href{mailto:kiraly@cs.elte.hu}{kiraly@cs.elte.hu}%
}{%
\href{http://people.inf.elte.hu/kpeter}{P\'eter KOV\'ACS}%
}{%
\href{http://www.chemaxon.com}{ChemAxon Ltd.} and\\%
\href{http://www.inf.elte.hu/alal/}{Dept. of Algorithms and Applications}\\%
\href{http://www.elte.hu/en}{E\"otv\"os Lor\'and University}\\%
Budapest, Hungary%
}{%
\href{mailto:kpeter@inf.elte.hu}{kpeter@inf.elte.hu}%
}

\short{%
Z. Kir\'aly, P. Kov\'acs
}{%
Efficient implementations of minimum-cost flow algorithms
}

\hyphenation{LEMON}
\hyphenation{LEDA}
\hyphenation{RelaxIV}
\hyphenation{MCFZIB}
\hyphenation{NETGEN}
\hyphenation{GOTO}
\hyphenation{supply}
\hyphenation{pseudo-flow}
\hyphenation{cycle}
\hyphenation{cycles}

\newcommand{\dash}{\nobreakdash-\hspace{0pt}}
\newcommand{\err}{\emph{error}}
\newcommand{\x}[1]{\textbf{#1}}
\newcommand{\pow}[2]{$\textrm{#1}^\textrm{#2}$}

\begin{abstract}
This paper presents efficient implementations of several algorithms
for solving the minimum-cost network flow problem.
Various practical heuristics and other
important implementation aspects are also discussed.
A~novel result of this work is the application of Goldberg's recent partial
augment-relabel method in the cost-scaling algorithm.
The presented implementations are available as part of the LEMON open source C++
optimization library (\url{http://lemon.cs.elte.hu/}).
The performance of these codes is compared to well-known and efficient
minimum-cost flow solvers, namely CS2, RelaxIV, MCF, and the corresponding
method of the LEDA library.
According to thorough experimental analysis, the presented cost-scaling
and network simplex implementations turned out to be more efficient than
LEDA and MCF.
Furthermore, the cost-scaling implementation is competitive with CS2.
The RelaxIV algorithm is often much slower than the other codes,
although it is quite efficient on particular problem instances.
\end{abstract}

\section{Introduction}
\label{sec:intro}
Network flow theory comprises a wide range of optimization models, which have
countless applications in various fields. One of the most fundamental
network flow problems is the \emph{minimum-cost flow} (MCF) problem.
It seeks a minimum-cost transportation of a specified amount of a commodity
from a set of supply nodes to a set of demand nodes in a directed network
with capacity constraints and linear cost functions defined on the arcs.
This problem directly arises in various real-world applications in
the fields of transportation, logistics, telecommunication, network design,
resource planning, scheduling, and many other industries.
Moreover, it also arises as a subproblem in more complex optimization tasks,
such as multicommodity flow problems.
For a comprehensive study of the theory, algorithms, and applications of
network flows, see the book of Ahuja, Magnanti, and Orlin
\cite{AMO93NetworkFlows}.

The MCF problem and its solution methods have been the object of
intensive research for decades and they have enormous literature.
Numerous algorithms have been developed and studied both from
theoretical and practical aspects
(see the books \cite{FordFulkerson62Flows, BusackerSaaty65Networks,
KenningtonHelgason80NetworkProg, AMO93NetworkFlows,
Schrijver03CombOpt, KorteVygen12CombOpt}).
Efficient implementation and profound experimental analysis of these
algorithms are also of high interest to the operations research community
(for example, see \cite{BradleyEtAl77Design, Grigoriadis86EfficientImpl,
BertsekasTseng94RelaxIV, Lobel96NetworkSimplex,
Goldberg97EfficientImpl, BunnagelEtAl98EfficientImpl, FrangioniManca06CostReopt}).
Nowadays, several commercial and non-commercial MCF solvers are available
under different license terms.

The primary goal of our research is to provide highly efficient
and robust open source implementations of different MCF algorithms
and to compare their performance in practice.
Preliminary work was published in \cite{KiralyKovacs10ExperimentalStudy}.
This paper presents a more detailed discussion of our implementations
along with extensive benchmark testing on a wide range of problem instances.

In order to achieve a comprehensive study, the following algorithms
were implemented:
SCC: a \emph{simple cycle-canceling} algorithm;
MMCC: \emph{minimum-mean cycle-canceling} algorithm;
CAT: \emph{cancel-and-tighten} algorithm;
SSP: \emph{successive shortest path} algorithm;
CAS: \emph{capacity-scaling} algorithm;
COS: \emph{cost-scaling} algorithm in three different variants; and
NS: primal \emph{network simplex} method with five different
pivot strategies.
All of these methods are generally known and well-studied algorithms, our
contribution is their efficient implementation with some new heuristics
and practical considerations.

According to the authors knowledge, our implementation of the
\emph{cost-scaling} algorithm is the first to apply
Goldberg's recent \emph{partial augment-relabel} method, which
was originally developed to solve the maximum flow problem efficiently
\cite{Goldberg08PartialAug}.
Its utilization in MCF algorithms was also suggested, but was not
investigated by Goldberg.
According to our tests, this new idea turned out to be a considerable
improvement in the cost-scaling MCF algorithm similarly to Goldberg's results
on the maximum flow algorithm.

This paper also presents an empirical evaluation of our implementations
and their variants.
Numerous benchmark tests were performed on many kinds of large-scale
networks containing up to millions of nodes and arcs.
These problem instances were created either using well-known random
generators, namely NETGEN and GOTO, or based on networks arising
in real-life problems.
The presented results demonstrate the relative performance of the
solution methods and give some guidelines for selecting an MCF algorithm
that is suitable for a desired application domain.

Our fastest implementations were also compared to four highly regarded
minimum-cost flow solvers:
CS2 code of Goldberg and Cherkassky \cite{Goldberg97EfficientImpl, CS2WEB},
an efficient authoritative implementation of the cost-scaling push-relabel
algorithm that has served as a benchmark for a long time;
the LEDA library~\cite{LEDA}, which also implements the cost-scaling algorithm;
L\"obel's MCF code \cite{Lobel96NetworkSimplex, LobelMCF}, which implements
the network simplex algorithm;
and RelaxIV \cite{MCFCLASS}, a C++ translation of the authoritative FORTRAN
implementation of the relaxation algorithm due to
Bertsekas and Tseng \cite{BertsekasTseng94RelaxIV, BertsekasNOC}.
We henceforth refer to the MCF code as MCFZIB in order to differentiate it
from the problem itself.
The experiments we conducted show that our cost-scaling implementation is
more efficient than LEDA and performs similarly to or slightly slower than CS2.
Our network simplex code clearly outperforms MCFZIB and it is the fastest
implementation for solving relatively small problem instances
(up to a few thousands of nodes).
For large networks, however, the cost-scaling codes are usually more efficient
than the network simplex algorithms.
The performance of RelaxIV turned out to be fluctuating: it is one of the
fastest implementations for solving certain kinds of problem instances, but
it is very slow for other instances.
The detailed experimental results can be found in Section~\ref{sec:testing}.

The implementations presented in this paper are available with full
source codes as part of the \href{http://lemon.cs.elte.hu}{LEMON} optimization
library \cite{LEMON}.
LEMON is an abbreviation of \emph{Library for Efficient Modeling and
Optimization in Networks}. It is an open source C++ template library
with focus on combinatorial optimization tasks related mainly to
graphs and networks.
It provides easy-to-use and highly efficient implementations of
graph algorithms and related data structures, which help solving
complex real-life optimization problems.
The LEMON project is maintained by the
\href{http://www.cs.elte.hu/egres}{MTA-ELTE Egerv\'ary Research Group
on Combinatorial Optimization (EGRES)} \cite{EGRES} at the
\href{http://www.cs.elte.hu/index.html?lang=en}{Department of Operations Research},
\href{http://www.elte.hu/en}{E\"otv\"os Lor\'and University},
Budapest, Hungary.
The library is also a member of the \href{http://www.coin-or.org}{COIN-OR}
initiative \cite{COINOR},
a collection of open source projects related to operations research.
LEMON applies a very permissive licensing scheme that makes it
favorable for commercial and non-commercial software development
as well as for research activities.
For more information about LEMON, the readers are referred to
the introductory paper \cite{DezsoJuttnerKovacs11Lemon} and to the
web site of the library: \url{http://lemon.cs.elte.hu/}.

The rest of this paper is organized as follows.
Section~\ref{sec:mcf} briefly introduces the MCF problem along with
the used notations and theorems.
Section~\ref{sec:algorithms} describes the implemented algorithms and their variants.
Section~\ref{sec:testing} presents the main experimental results.
Finally, the conclusions are drawn in Section~\ref{sec:conclusions}.

\section{The minimum-cost flow problem}
\label{sec:mcf}

\subsection{Definitions and notations}
\label{sec:definitions}
The \emph{minimum-cost flow} (MCF) problem is defined as follows.
Let $G=(V,A)$ be a weakly connected directed graph consisting of
$n=|V|$ nodes and $m=|A|$ arcs.
We associate with each arc $(i,j)\in A$
a \emph{capacity} (upper bound) $u_{ij}\geq 0$
and a \emph{cost} $c_{ij}$, which denotes the cost per unit flow on the arc.
Each node $i\in V$ has a signed \emph{supply} value $b_i$.
If $b_i > 0$, then node $i$ is called a \emph{supply node} with a
supply of $b_i$; if $b_i < 0$, then node $i$ is called a \emph{demand node}
with a demand of $-b_i$; and if $b_i = 0$, then node $i$ is
referred to as a \emph{transshipment node}.
We assume that all data are integer and we wish to find an integer-valued
flow of minimum total cost satisfying the supply-demand constraints
at all nodes and the capacity constraints on all arcs.
The solution of the problem is represented by flow
values~$x_{ij}$ assigned to the arcs.
Therefore, the MCF problem can be stated as
\begin{subequations}
\begin{equation}
  \label{eq:mcf1}
  \min\;\; \sum_{(i,j)\in A} c_{ij}x_{ij}
\end{equation}
subject to
\begin{equation}
  \label{eq:mcf2}
  \sum_{j:(i,j)\in A} x_{ij} - \sum_{j:(j,i)\in A} x_{ji} = b_i
    \qquad  \forall i\in V\textrm{,}
\end{equation}
\begin{equation}
  \label{eq:mcf3}
  0\leq x_{ij}\leq u_{ij} \qquad \forall (i,j)\in A\textrm{.}
\end{equation}
\end{subequations}

We refer to \eqref{eq:mcf2} as \emph{flow conservation constraints}
and \eqref{eq:mcf3} as \emph{capacity constraints}.
A solution vector $x$ is called \emph{feasible} if it satisfies all
constraints defined in \eqref{eq:mcf2} and \eqref{eq:mcf3},
and it is called \emph{optimal} if it also minimizes
the total flow cost \eqref{eq:mcf1} over the feasible solutions.

The flow conservation constraints \eqref{eq:mcf2}
imply that the sum of the node supply values
is required to be zero, that is, $\sum_{i\in V} b_i = 0$,
in order to have a feasible solution to the MCF problem:
\begin{equation*}
  \sum_{i\in V} b_i
    = \sum_{i\in V} \Big( \sum_{j:(i,j)\in A} x_{ij} -
      \sum_{j:(j,i)\in A} x_{ji} \Big)
    = \sum_{i\in V} \sum_{j:(i,j)\in A} x_{ij} -
      \sum_{i\in V} \sum_{j:(j,i)\in A} x_{ji} = 0\textrm{.}
\end{equation*}
Without loss of generality, we may further assume that
all arc capacities are finite, all arc costs are nonnegative,
and the problem has a feasible solution~\cite{AMO93NetworkFlows}.

There are several other problem formulations that are equivalent
to the above definition,
for instance, the minimum-cost circulation problem, the uncapacitated
minimum-cost flow problem, and the transportation problem.
However, this definition is quite common in the literature.

Flow algorithms and related theorems usually rely on
the concept of \emph{residual networks}
\cite{FordFulkerson62Flows, BusackerSaaty65Networks,
AMO93NetworkFlows, Schrijver03CombOpt, KorteVygen12CombOpt}.
For a given feasible flow~$x$, the corresponding
residual network $G_x$ is defined as follows.
Let $G_x = (V,A_x)$ be a directed graph that contains
\emph{forward} and \emph{backward arcs} on the original
node set $V$.
A forward arc $(i,j)\in A_x$ corresponds to each original arc
$(i,j)\in A$ for which the \emph{residual capacity}
$r_{ij} = u_{ij} - x_{ij}$ is positive.
A backward arc $(j,i)\in A_x$ corresponds to each original arc
$(i,j)\in A$ for which the \emph{residual capacity}
$r_{ji} = x_{ij}$ is positive.
The cost of a forward arc $(i,j)$ is defined as $c_{ij}$,
while the cost of a backward arc $(j,i)$ is $-c_{ij}$.

The concept of \emph{pseudoflows} is also important for several
flow algorithms.
A~pseudoflow is a function~$x$ defined on the arcs that satisfies only the
nonnegativity and capacity constraints~\eqref{eq:mcf3} but might violate
the flow conservation constraints~\eqref{eq:mcf2}.
A feasible flow is also a pseudoflow.
In case of a pseudoflow~$x$, a node might have a certain amount of undelivered
supply or unfulfilled demand, which is called the \emph{excess} or
\emph{deficit} of the node, respectively.
Formally, the signed \emph{excess} value of a node $i$ with respect to a
pseudoflow~$x$ is defined as
\begin{equation}
  \label{def:excess}
  e_i = b_i + \sum_{j:(j,i)\in A} x_{ji} - \sum_{j:(i,j)\in A} x_{ij}\textrm{.}
\end{equation}
If $e_i > 0$, node $i$ is referred to as an \emph{excess node} with an
excess of $e_i$; and if $e_i < 0$, node $i$ is called a \emph{deficit node}
with a deficit of $-e_i$.
Note that $\sum_{i\in V} e_i = \sum_{i\in V} b_i = 0$, that is,
the total excess of the nodes equals to the total deficit.
The residual network corresponding to a pseudoflow is defined in the same
way as in case of a feasible flow.

The running time of an MCF algorithm is measured as a function of
the size of the network and the magnitudes of the input data.
Let $U$ henceforth denote the largest node supply or arc capacity:
\begin{equation}
  \label{def:maxu}
  U = \max\{\max\{ |b_{i}| : i\in V\}, \max\{ u_{ij} : (i,j)\in A\} \}
\end{equation}
and let $C$ denote the largest arc cost:
\begin{equation}
  \label{def:maxc}
  C = \max\{ c_{ij} : (i,j)\in A\} \textrm{.}
\end{equation}
An algorithm is referred to as \emph{pseudo-polynomial} if its running
time is bounded by a polynomial function in the dimensions of the problem and
the magnitudes of the numerical data, namely, $n$, $m$, $U$, and $C$.
These algorithms technically run in exponential time with respect to the size of
the input and they are therefore not considered polynomial.
A \emph{(weakly) polynomial} algorithm is one that runs in time polynomial in the
input size, namely, $n$, $m$, $\log U$, and $\log C$.
Furthermore, an algorithm is called \emph{strongly polynomial} if its
running time depends upon only on the inherent dimensions of the problem, that is,
it runs in time polynomial in $n$ and $m$ regardless of the numerical input data.

\subsection{Optimality conditions}
\label{sec:optimality}
In the followings, we formulate optimality conditions for the MCF
problem in terms of the residual network as well as the original network.
These fundamental theorems are useful in several aspects.
Not only do they provide simple methods for verifying the
optimality of a certain solution, but they also suggest
algorithms for solving the problem.
These results are discussed in
\cite{FordFulkerson62Flows, BusackerSaaty65Networks,
AMO93NetworkFlows, Schrijver03CombOpt, KorteVygen12CombOpt}.

\begin{theorem}[Negative cycle optimality conditions]
\label{thm:negcyc_optimality}
  A feasible solution $x$ of the MCF problem is optimal
  if and only if the residual network $G_x$ contains no directed cycle
  of negative total cost.
\end{theorem}

This theorem is a consequence of the observation that
any feasible flow can be decomposed into a finite set of
augmenting paths and cycles.

We also introduce two equivalent formulations of optimality conditions that
rely on the notions of \emph{node potentials} and \emph{reduced costs}.
We associate with each node $i\in V$ a signed value $\pi_i$,
which is referred to as the potential of node $i$.
Actually, $\pi_i$ can be viewed as the linear programming dual
variable corresponding to the flow conservation constraint
of node $i$ (see \cite{AMO93NetworkFlows}).
With respect to a given potential function $\pi$,
the reduced cost of an arc $(i,j)$ is defined as
\begin{equation}
  \label{def:reducedcost}
  c_{ij}^\pi = c_{ij} + \pi_i - \pi_j\textrm{.}
\end{equation}
Note that $c_{ij}^\pi$ measures the relative cost of the arc $(i,j)$
with respect to the potentials of its end-nodes.

This concept allows us to formulate the following optimality conditions.

\begin{theorem}[Reduced cost optimality conditions]
\label{thm:redcost_optimality}
  A feasible solution~$x$ of the MCF problem is optimal
  if and only if for some node potential function~$\pi$, the
  reduced cost of each arc in the residual network $G_x$
  is nonnegative:
  \begin{equation}
    c_{ij}^\pi \geq 0 \qquad \forall (i,j)\in A_x\textrm{.}
  \end{equation}
\end{theorem}

Note that the total reduced cost of a directed cycle with respect to
any potential function equals to the original cost of the cycle.
Therefore, the conditions of Theorem~\ref{thm:redcost_optimality}
obviously imply the negative cycle optimality conditions
defined in Theorem~\ref{thm:negcyc_optimality}.
Furthermore, a constructive proof exists for the converse result.
For an optimal flow~$x$, corresponding optimal node potentials~$\pi$
can be obtained by solving a shortest path problem
in the residual network.

Theorem~\ref{thm:redcost_optimality} can be restated in terms of the original
network as follows.

\begin{theorem}[Complementary slackness optimality conditions]
\label{thm:compslack_optimality}
  A feasible solution $x$ of the MCF problem is optimal
  if and only if for some node potential function~$\pi$,
  the following complementary slackness conditions hold
  for each arc $(i,j)\in A$ of the original network:
  \begin{subequations}
  \begin{align}
    & \textrm{if } c_{ij}^\pi > 0\textrm{, then }
      x_{ij} = 0\textrm{;}\\
    & \textrm{if } 0 < x_{ij} < u_{ij}\textrm{, then }
      c_{ij}^\pi  = 0\textrm{;}\\
    & \textrm{if } c_{ij}^\pi < 0\textrm{, then }
      x_{ij} = u_{ij}\textrm{.}
  \end{align}
  \end{subequations}
\end{theorem}

In addition to these exact optimality conditions, the characterization
of \emph{approximate optimality} is also of particular importance.
Several algorithms rely on the concept of $\epsilon$\dash optimality.
For a given $\epsilon\geq 0$, a feasible flow or a pseudoflow~$x$
is called \emph{$\epsilon$\dash optimal} if for some node potential function~$\pi$,
the reduced cost of each arc in the residual network $G_x$ is at least
$-\epsilon$, that is,
\begin{equation}
  c_{ij}^\pi \geq -\epsilon \qquad \forall (i,j)\in A_x\textrm{.}
\end{equation}
These conditions are the relaxations of the reduced cost
optimality conditions defined in Theorem~\ref{thm:redcost_optimality} and are
equivalent to them when $\epsilon=0$.
The $\epsilon$\dash optimality conditions can also be restated in terms
of the original network to obtain the relaxations of the
complementary slackness optimality conditions defined in
Theorem~\ref{thm:compslack_optimality}.

The following lemma formulates two simple observations that are
related to $\epsilon$\dash optimality.

\begin{lemma}
\label{thm:epsopt_properties}
  Any feasible solution $x$ of the MCF problem is $\epsilon$\dash optimal
  if $\epsilon\geq C$.
  Moreover, if the arc cost are integer and $\epsilon < 1/n$,
  then an $\epsilon$\dash optimal feasible flow is an optimal solution.
\end{lemma}

Note that an $\epsilon$\dash optimal flow is also $\epsilon'$\dash optimal for
all $\epsilon'>\epsilon$ and hence the approximate optimality of $x$
is best indicated by the smallest value $\epsilon\geq 0$ for which $x$
is $\epsilon$\dash optimal.
This minimum value is referred to as $\epsilon(x)$.
The following theorem reveals an inherent connection between $\epsilon(x)$
and the \emph{minimum-mean cycles} of the residual network.
The \emph{mean cost} of a directed cycle is defined as its total cost
divided by the number of arcs in the cycle.

\begin{theorem}
\label{thm:epsopt_mmc}
  For a non-optimal feasible solution $x$ of the MCF problem, $\epsilon(x)$
  equals to the negative of the minimum-mean cost of a directed cycle in
  the residual network $G_x$.
  For an optimal solution $x$, $\epsilon(x)=0$.
\end{theorem}

Therefore, $\epsilon(x)$ can be computed by finding a directed cycle of
minimum-mean cost, which can be carried out in $O(nm)$
time \cite{Karp78MinCycleMean}.
Another related problem is to find an appropriate potential function~$\pi$
for an $\epsilon$\dash optimal flow or pseudoflow~$x$ so that they satisfy the
$\epsilon$\dash optimality conditions.
Similarly to the problem of finding optimal node potentials, this problem
can also be solved by performing a shortest path computation
in $G_x$ but with a modified cost function $c'$ for which
$c'_{ij} = c_{ij} + \epsilon$ for each arc $(i,j)$ in $G_x$.
See, for example, \cite{AMO93NetworkFlows, Frank11Connections} for
the proof of all these results related to $\epsilon$\dash optimality.

\subsection{Solution methods}
\label{sec:history}
The MCF problem and its solution methods have a rich history spanning
more than fifty years.
Researchers first studied a classical special case of the MCF problem, the so-called
\emph{transportation problem},
in which the network consists only of supply and demand nodes.
Dantzig was the first to solve the
transportation problem by specializing his famous linear programming method,
the simplex algorithm.
Later, he also applied this approach to the MCF problem and developed a solution
method that is known as the \emph{network simplex} algorithm.
These results are discussed in Dantzig's book \cite{Dantzig63LP}.

Ford and Fulkerson
developed the first combinatorial algorithms for the uncapacitated and
capacitated transportation problems
by generalizing Kuhn's remarkable Hungarian Method \cite{Kuhn55HungarianMethod}.
Ford and Fulkerson later proposed a similar \emph{primal--dual} algorithm for
the MCF problem, as well.
Their results are presented in the book \cite{FordFulkerson62Flows}.

In the next few years, other algorithmic approaches were also suggested,
namely, the \emph{successive shortest path} algorithm, the \emph{out-of-kilter}
algorithm, and the \emph{cycle-canceling} algorithm.
These methods, however, do not run in polynomial time.
Therefore, both theoretical and practical expectations motivated further
research on developing more efficient algorithms.
Edmonds and Karp \cite{EdmondsKarp72Theoretical} introduced the
\emph{scaling technique} and developed the first weakly polynomial-time
algorithm for solving the MCF problem.
Later, other researchers also recognized the significant value of this
approach and proposed various scaling algorithms.
The problem of finding a strongly polynomial-time MCF algorithm, however,
remained a challenging open question for several years.
Tardos \cite{Tardos85StronglyPoly} developed the first such algorithm, which was
followed by many other methods providing improved running time bounds.

Besides theoretical aspects, efficient implementation and computational
evaluation of MCF algorithms have also been an object of intensive research.
The \emph{network simplex} algorithm became quite popular in practice
when efficient spanning tree labeling techniques were developed to
improve its performance.
Later, other algorithms also turned out to be quite efficient.
Implementations of \emph{relaxation} and \emph{cost-scaling} algorithms were
reported to be competitive with the fastest network simplex codes.

Detailed discussion and complexity survey of MCF algorithms
can be found in, for example, \cite{AMO93NetworkFlows,
Schrijver03CombOpt, KorteVygen12CombOpt}.
Table~\ref{tab:complexity} provides a brief summary of the MCF algorithms
having best theoretical running time.
Recall from Section~\ref{sec:definitions} that $n$ and $m$ denote the number of
nodes and arcs in the network, respectively; $U$~denotes the
maximum of supply values and arc capacities; and $C$ denotes the
largest arc cost.
Furthermore, let $\textrm{SP}_+(n,m)$ denote the running time of any
algorithm solving the single-source shortest path problem in a directed
graph with $n$ nodes, $m$ arcs, and a nonnegative length function.
Dijkstra's algorithm with Fibonacci heaps provides an $O(m + n\log n)$
bound for $\textrm{SP}_+(n,m)$
\cite{Schrijver03CombOpt, CLRS09Algorithms, KorteVygen12CombOpt}.

\begin{table}[!htb]
  \centering
  \footnotesize
  \begin{tabular}{|c|l|}
    \hline & \\[-2.25ex]
    \multirow{2}{*}{$O(nU\cdot\textrm{SP}_+(n,m))$} &
      Edmonds and Karp \cite{EdmondsKarp72Theoretical};
      Tomizawa \cite{Tomizawa71Techniques}
      \\ & \emph{successive shortest path}
    \\[0.25ex]\hline & \\[-2.25ex]
    \multirow{2}{*}{$O(m\log U\cdot\textrm{SP}_+(n,m))$} &
      Edmonds and Karp \cite{EdmondsKarp72Theoretical}
      \\ & \emph{capacity-scaling}
    \\[0.25ex]\hline & \\[-2.25ex]
    \multirow{2}{*}{$O(m\log n\cdot\textrm{SP}_+(n,m))$} &
      Orlin \cite{Orlin93Faster}
      \\ & \emph{enhanced capacity-scaling}
    \\[0.25ex]\hline & \\[-2.25ex]
    \multirow{2}{*}{$O(nm\log (n^2/m)\log(nC))$} &
      Goldberg and Tarjan \cite{GoldbergTarjan90SuccApprox}
      \\ & \emph{generalized cost-scaling}
    \\[0.25ex]\hline & \\[-2.25ex]
    \multirow{2}{*}{$O(nm\log\log U\log(nC))$} &
      Ahuja, Goldberg, Orlin, and Tarjan \cite{AhujaEtAl92DoubleScaling}
      \\ & \emph{double scaling}
    \\[0.25ex]\hline & \\[-2.25ex]
    \multirow{1}{*}{$O((m^{3/2}U^{1/2} + mU\log(mU))\log(nC))$} &
      Gabow and Tarjan \cite{GabowTarjan89FasterScaling}
    \\[0.25ex]\hline & \\[-2.25ex]
    \multirow{1}{*}{$O((nm + mU\log(mU))\log(nC))$} &
      Gabow and Tarjan \cite{GabowTarjan89FasterScaling}
    \\[0.25ex]\hline
  \end{tabular}
  \caption{Best theoretical running time bounds for the MCF problem}
  \label{tab:complexity}
\end{table}

\section{Implemented algorithms}
\label{sec:algorithms}
This section discusses the algorithms we implemented as well as the most
important heuristics and other practical improvements.
All of these methods are well-studied in the literature.
Their profound theoretical analysis with the proof of
correctness and running time can be found in \cite{AMO93NetworkFlows}
and in other papers and books cited later in this section.
The contribution of this paper is the efficient implementation and
empirical analysis of several variants of these algorithms.
For further details of our implementations, the readers are referred to
the documentation and the source code of the LEMON
library \cite{LEMON}.

Table~\ref{tab:algorithms} provides an overview of the
implemented algorithms and their worst-case running time.
The same notations are used as in the previous section.
Two of these algorithms perform shortest path computations with
nonnegative length functions.
Our implementations use Dijkstra's algorithm with binary heaps
by default, hence $\textrm{SP}_+(n,m) = O((n + m)\log n)$.
Note that some of these algorithms have other variants with better
theoretical running time, but our research especially focused on
the practical performance of them.
The given running time bounds correspond to the actual implementations.

\begin{table}[!htb]
  \centering\small
  \begin{tabular}{lll}
    \hline & & \\[-10pt]
    Alg. & Name & Running time\\[1pt]
    \hline & & \\[-7pt]
    SCC & \emph{simple cycle-canceling}
      & $O(nm^2CU)$\\[1pt]
    MMCC & \emph{minimum-mean cycle-canceling}
      & $O(n^2m^2\min\{\log(nC), m\log n\})$ \\[1pt]
    CAT & \emph{cancel-and-tighten}
      & $O(n^2m\min\{\log(nC), m\log n\})$ \\[1pt]
    SSP & \emph{successive shortest path}
      & $O(nU\cdot\textrm{SP}_+(n,m))$ \\[1pt]
    CAS & \emph{capacity-scaling}
      & $O(m\log U\cdot\textrm{SP}_+(n,m))$ \\[1pt]
    COS & \emph{cost-scaling}
      & $O(n^2m \log(nC))$ \\[1pt]
    NS & \emph{network simplex}
      & $O(nm^2CU)$ \\[2pt]
    \hline
  \end{tabular}
  \caption{Implemented algorithms and their worst-case running time}
  \label{tab:algorithms}
\end{table}

\subsection{Cycle-canceling algorithms}
\label{sec:cyclecanceling}

Cycle-canceling is one of the simplest methods for solving the MCF problem.
This algorithm applies a primal approach based on
Theorem~\ref{thm:negcyc_optimality}. A feasible flow~$x$ is first established,
which can be carried out by solving a maximum flow problem.
After that, the algorithm throughout maintains feasibility of the solution~$x$
and gradually decreases its total cost.
At each iteration, a directed cycle of negative cost is identified in
the residual network~$G_x$ and this cycle is canceled by pushing
the maximum possible amount of flow along it.
When the residual network contains no negative-cost directed cycle,
the algorithm terminates and Theorem~\ref{thm:negcyc_optimality}
implies that the solution is optimal.

The cycle-canceling algorithm was proposed by Klein \cite{Klein67PrimalMethod}.
Its generic version does not specify the order of selecting negative cycles
to be canceled, but it runs in pseudo-polynomial time for the MCF problem with
integer data.
Since the total flow cost is decreased at each iteration and
$mCU$ is clearly an upper bound of the flow cost, the algorithm
performs $O(mCU)$ iterations if all data are integer.
Klein used a label-correcting shortest path algorithm that identifies
a negative cycle in $O(nm)$ time, thus his algorithm runs in
$O(nm^2CU)$ time.
Later, numerous other variants of the cycle-canceling method were also
developed by applying different rules for cycle selection, for example
\cite{BarahonaTardos89Note, GoldbergTarjan89CancelingCycles,
SokkalingamEtAl00NewPoly}.
These algorithms have quite different theoretical and practical
behavior. Some of them run in polynomial or even
strongly polynomial time.

We implemented three cycle-canceling algorithms,
which are discussed in the followings.

\paragraph{Simple cycle-canceling algorithm.}
This is a simple version of the cycle-canceling method
using the Bellman--Ford algorithm for identifying negative cycles.
We henceforth denote this implementation as SCC.

It is well-known that the Bellman--Ford algorithm is capable of detecting
a negative-cost directed cycle after performing $n$ iterations
or detecting that such a cycle does not exist \cite{CLRS09Algorithms}.
However, it is not required to perform $n$ iterations in most cases.
If negative cycles exist in the graph, one or more of them
typically appear in the subgraph identified by the predecessor
pointers of the nodes after much less iterations.
Unfortunately, we do not know the sufficient limit for the number of
iterations in advance and searching for cycles using the
predecessor pointers at an intermediate step of the algorithm is a
relatively slow operation.
Therefore, our SCC implementation performs such checking after a
successively increasing number of iterations of the Bellman--Ford algorithm.
According to our tests, it turned out to be practical to search for
negative cycles after executing $\lfloor 2\cdot1.5^k \rfloor$
iterations for each $k \geq 0$ until this limit reaches $n$.
It is also beneficial to cancel all node-disjoint negative cycles
that can be found at once when Bellman--Ford algorithm is stopped.
The worst case time complexity of the SCC algorithm is $O(nm^2CU)$.

\paragraph{Minimum-mean cycle-canceling algorithm.}
This famous special case of the cycle-canceling method was developed
by Goldberg and Tarjan \cite{GoldbergTarjan89CancelingCycles}. It selects
a negative cycle of minimum mean cost to be canceled at each iteration,
which yields the simplest MCF algorithm running in strongly polynomial time.
We denote this method and our implementation as MMCC.

Recall from Section~\ref{sec:optimality} that the mean cost of a
directed cycle is defined as its total cost divided by the
number of arcs in the cycle.
The MMCC algorithm iteratively identifies a directed cycle~$W$ of minimum-mean
cost in the current residual network.
If the cost of $W$ is negative, then the cycle is canceled
and another iteration is performed, otherwise the algorithm terminates
with an optimal solution found.
It has been proved that this algorithm performs $O(nm^2 \log n)$
iterations for arbitrary real-valued arc costs and
$O(nm \log(nC))$ iterations for integer arc costs.
The proof of these bounds relies on the concept of $\epsilon$\dash optimality
and is rather involved, see
\cite{GoldbergTarjan89CancelingCycles, AMO93NetworkFlows, KorteVygen12CombOpt},
although the algorithm is very simple to state.

The MMCC algorithm relies on finding minimum-mean directed cycles in a graph.
This optimization problem has also been studied for a long time and
several efficient algorithms have been developed for solving it
\cite{Karp78MinCycleMean, HartmannOrlin93Finding,
DasdanGupta98FasterAlg, Dasdan04Experimental,
GeorgiadisEtAl09Experimental}.
The best strongly polynomial-time bound for a minimum-mean cycle
algorithm is $O(nm)$ and thus the overall running time of the MMCC algorithm
is $O(n^2m^2\min\{\log(nC), m\log n\})$ for the MCF problem with
integer data.

We implemented three known algorithms for finding minimum-mean cycles:
Karp's original algorithm \cite{Karp78MinCycleMean};
an improved version of this method that is due to Hartmann and Orlin
\cite{HartmannOrlin93Finding};
and Howard's \emph{policy-iteration} algorithm
\cite{Howard60DynamicProg, Dasdan04Experimental}.
The first two methods run in strongly polynomial time $O(nm)$.
In contrast, Howard's algorithm is not known to be polynomial,
but it is one of the fastest solution methods in
practice \cite{DasdanGupta98FasterAlg, Dasdan04Experimental}.

Our experiments also verified that Howard's algorithm is orders of magnitude
faster than the other two methods we implemented.
This algorithm gradually approximates the optimal solution by
performing linear-time iterations.
Relatively few iterations are typically sufficient to find a minimum-mean
cycle, but no polynomial upper bound is known.
Therefore, we developed a combined method in order to achieve the best
performance in practice while keeping the strongly polynomial upper bound
on the running time.
Howard's algorithm is run with an explicit limit on the number of
iterations. If this limit is reached without finding the optimal solution,
we stop Howard's algorithm and execute the Hartmann--Orlin algorithm.
We set this iteration limit to $n$, and hence the overall running time of
this combined method is $O(nm)$, which equals to the best strongly
polynomial bound.
In our experiments, the iteration limit was indeed never reached.
Thus, the combined method was practically identical to Howard's
algorithm but with a guarantee of worst-case running time $O(nm)$.

\paragraph{Cancel-and-tighten algorithm.}
This algorithm can be viewed as an improved version of the MMCC algorithm, which
is also due to Goldberg and Tarjan \cite{GoldbergTarjan89CancelingCycles}.
It is faster than MMCC both in theory and practice.
This algorithm is henceforth denoted as CAT.

The improvement of the CAT algorithm is based on
a more flexible selection of cycles to be canceled.
The previously studied cycle-canceling algorithms are pure primal
methods in a sense that they do not consider the dual solution at all.
In contrast, the CAT algorithm explicitly maintains
node potentials, which make the detection of negative residual
cycles easier and faster.
The key idea of the algorithm is to cancel cycles that consist
entirely of negative-cost arcs.
Note that the sum of the reduced arc costs along a cycle with respect to
any potential function is exactly the same as the original cost of the cycle.
Therefore, the algorithm can consider the reduced costs with respect to
the current node potentials instead of the original costs.
A residual arc is called \emph{admissible} if its reduced cost is negative;
the subgraph of the residual network consisting only of the admissible arcs
is called the \emph{admissible network}; and a directed cycle
in the admissible network is referred to as an \emph{admissible cycle}.

The CAT algorithm performs two main steps at every iteration until the
current solution becomes optimal.
In the \emph{cancel} step, admissible cycles are successively canceled
until such a cycle does not exist.
In the \emph{tighten} step, the node potentials are modified
in order to make more arcs admissible.
Despite the MMCC algorithm, this method explicitly utilizes the
concept of $\epsilon$\dash optimality.
Recall the corresponding definitions and theorems from
Section~\ref{sec:optimality}.
The CAT algorithm ensures $\epsilon$\dash optimality of the solution
for successively smaller values of $\epsilon\geq 0$.
In the tighten step, the potentials are modified so as to satisfy
the $\epsilon$\dash optimality conditions for a smaller $\epsilon$
that is at most $(1-1/n)$ times its former value.

The cancel step is the dominant part of the computation.
We implemented a straightforward method for this step based on
a depth-first traversal of the admissible network.
This implementation runs in $O(nm)$ time as canceling a cycle
takes $O(n)$ time and at most $O(m)$ admissible cycles can be successively
canceled without modifying the potential function.
Goldberg and Tarjan \cite{GoldbergTarjan89CancelingCycles} also showed that
using dynamic tree data structures \cite{SleatorTarjan83DynamicTree},
the running time of this step can be reduced to $O(m\log n)$
(amortized time $O(\log n)$ per cycle cancellation).
However, we did not invest effort in implementing this variant
because the cycle-canceling algorithms turned out to be relatively
slow in our experimental tests (see Section~\ref{sec:testing}).

The tighten step can be performed in $O(m)$ time based on a
topological ordering of the nodes with respect to the
admissible network.
This implementation, however, does not ensure that the overall
running time of the algorithm is strongly polynomial.
To overcome this drawback,
Goldberg and Tarjan \cite{GoldbergTarjan89CancelingCycles}
suggested to carry out the tighten step in a stricter way
after every $O(n)$ iterations of the algorithm.
In these cases, a minimum-mean cycle computation is performed
to exactly determine the smallest $\epsilon$ value
for which the current flow is $\epsilon$\dash optimal
(see Theorem~\ref{thm:epsopt_mmc}).
Node potentials are also recomputed to correspond to this
$\epsilon$ value.
Note that the amortized running time $O(m)$ of the tighten step is
not affected by this modification.
Our implementation, however, performs this stricter tighten step more often,
namely after every $\lfloor\sqrt{n}\rfloor$ iterations, because it
turned out to be more efficient in practice.
This means that the amortized running time of the tighten step becomes
$O(m\sqrt{n})$, but it is still less than the $O(nm)$ time of our
implementation of the cancel step.
The minimum-mean cycle computations are carried out using the same
combined algorithm that was applied in the MMCC algorithm.

This algorithm is strongly polynomial.
It runs in $O(n^2m^2\log n)$ time for the MCF problem with
arbitrary arc costs and in $O(n^2m \log(nC))$
time for integer arc costs, see \cite{GoldbergTarjan89CancelingCycles}.

\medskip
The experimental results for these algorithms are presented in
Section~\ref{sec:testing}.
It~turned out that their relative performance depends upon
the problem instance, but the CAT algorithm is usually much more
efficient than both SCC and MMCC.
However, all three of these cycle-canceling algorithms turned out to be
slower than the cost-scaling and network simplex methods.

\subsection{Augmenting path algorithms}
\label{sec:augpath}
Another fundamental approach for solving the MCF problem is the
so-called \emph{successive shortest path} method.
It is a dual ascent algorithm that successively augments flow along
shortest paths of the residual network to send the required amount of
flow from the supply nodes to the demand nodes.
In this sense, this method can be viewed as a generalization of the
well-known augmenting path algorithms solving the maximum flow problem,
namely the Ford--Fulkerson and Edmonds--Karp algorithms
\cite{FordFulkerson62Flows, EdmondsKarp72Theoretical,
AMO93NetworkFlows, CLRS09Algorithms}.

The successive shortest path algorithm in its inital form was developed
independently by
\mbox{Jewell} \cite{Jewell58Optimal}, Iri \cite{Iri60NewMethod}, and
Busacker and Gowen \cite{BusackerGowen60Procedure}.
They showed that the MCF problem can be solved by a sequence of shortest
path computations.
Later, Edmonds and Karp \cite{EdmondsKarp72Theoretical} and
Tomizawa \cite{Tomizawa71Techniques} independently suggested
the utilization of node potentials in the algorithm to maintain
nonnegative arc costs for the shortest path problems.
This technique greatly improves both the theoretical and the practical
performance of the algorithm.
Edmonds and Karp \cite{EdmondsKarp72Theoretical} also developed a
capacity-scaling variant of this method that runs in polynomial time.

We implemented the standard successive shortest path algorithm applying
node potentials as well as the capacity-scaling method.
These algorithms and the most important aspects of their implementations
are discussed below.

\paragraph{Successive shortest path algorithm.}
This algorithm is henceforth denoted as SSP.
In contrast to the cycle-canceling method, which maintains a feasible
flow and attempts to achieve optimality, the SSP algorithm maintains
an optimal pseudoflow and node potentials and attempts to achieve feasibility.
Recall from Section~\ref{sec:definitions} that a pseudoflow satisfies the
nonnegativity and capacity constraints but might violate the flow conservation
constraints at some nodes.
Such a node has a certain amount of excess or deficit.

The SSP algorithm begins with constant zero pseudoflow~$x$ and a constant
potential function~$\pi$ and proceeds by gradually converting $x$
into a feasible solution while throughout maintaining the reduced cost
optimality conditions defined in Theorem~\ref{thm:redcost_optimality}.
At every iteration, the algorithm selects a node with positive excess
and sends flow from this node to an arbitrary deficit node along a shortest
path of the residual network with respect to the reduced arc costs.
After that, the node potentials are modified using the computed
shortest path distances to preserve the reduced cost optimality conditions.
These conditions not only verify the optimality of both the
primal and the dual solutions, but they also ensure nonnegative
arc costs for the consecutive shortest path computations.
By sending flow from excess nodes to deficit nodes, the algorithm
iteratively decreases the total excess of the nodes
until the solution becomes feasible.
At the beginning of the algorithm, the total excess is
at most $nU/2$ and each iteration decreases this value by at least one
(in case of integer data), thus the SSP algorithm terminates after
$O(nU)$ path augmentations.
The flow conservation constraints are then satisfied at all nodes
and hence the solution is both feasible and optimal.

We implemented the SSP algorithm as follows.
At each iteration, flow is augmented from the current excess node~$v$
to a deficit node~$w$ whose shortest path distance from $v$ is minimal.
The shortest path searches are carried out using Dijkstra's algorithm
with a heap data structure.
We experimented with several heap variants provided by the LEMON
library \cite{LEMONDOC} and the standard binary heap structure turned out
to be one of the fastest and most robust implementations.
Therefore, our SSP implementation uses this data structure by default.
This means that a single iteration is performed in $O((n + m)\log n)$ time
and the overall complexity of the algorithm is $O(nU (n + m)\log n)$.
However, one can easily switch to other data structures
(for instance, Fibonacci heaps).

The practical performance of the SSP algorithm mainly depends on the
shortest path computations.
We applied a significant improvement related to these searches,
which is discussed, for example, in \cite{AMO93NetworkFlows}.
At each iteration, it is not necessary to compute the shortest paths
to all nodes from the current excess node $v$, but
the Dijkstra algorithm can be terminated once it permanently labels
a deficit node $w$.
The node potentials can also be updated in an alternative way that
does not require any modification for those nodes that were not
permanently labeled during the shortest path computation.
This improvement can be implemented quite easily, but it greatly improves
the efficiency of the SSP algorithm in practice.

The representation of the residual network is another important aspect
of the implementation.
It is possible to implement the SSP algorithm using the original representation
of the input network, but in this case, all outgoing and incoming arcs
of the current node have to be checked at each step of the shortest
path computations.
Another possibility is to explicitly maintain the residual network
containing only those arcs that have positive residual capacity.
However, this implementation would require the updating of the graph structure
after each path augmentations, which is time-consuming.

We applied an intermediate solution that turned out to be
the most efficient.
We store an auxiliary graph $G'$ that contains all possible forward and
backward arcs and also maintain their residual capacities explicitly.
All shortest path computations run on $G'$ by skipping those arcs whose
current residual capacity is zero.
The flow augmentations are carried out by decreasing the residual capacities
of the arcs on the path and increasing the residual capacities of the
corresponding reverse arcs.
Therefore, we also store for each arc an index to its \emph{reverse arc}
(often referred to as \emph{sister arc}).
The major benefit of this implementation is that this auxiliary graph $G'$ allows
a quite efficient representation.
Note that using $G'$, only the outgoing arcs of a node have to be
traversed during the shortest path searches and $G'$ is not modified
throughout the algorithm.
We can, therefore, represent the outgoing arcs of a node in $G'$
by consecutive integers, which makes it possible to traverse these arcs
quite efficiently without iterating over the elements of
an array or a linked list.

The reduced arc costs are also required in the shortest path computations.
Since these values are frequently modified by adjusting node potentials,
it is better to store only the potentials and
recompute reduced costs whenever they are needed.
Furthermore, we also maintain a signed excess value for each node.

\paragraph{Capacity-scaling algorithm.}
This algorithm, which we denote as CAS, is an improved version of the
SSP method. It uses a capacity-scaling scheme that reduces the number of
iterations from $O(nU)$ to $O(m\log U)$.
This algorithm was devised by Edmonds and Karp \cite{EdmondsKarp72Theoretical}
as the first weakly polynomial-time solution method for the MCF problem.
Our implementation is based on a slightly modified variant that is
due to Orlin \cite{Orlin93Faster} and also
discussed in \cite{AMO93NetworkFlows}.

The SSP algorithm has a substantial drawback that
the shortest path augmentations might deliver relatively small amounts of flow,
which results in a large number of iterations.
This is overcome in the CAS algorithm by ensuring that each path augmentation
carries a sufficiently large amount of flow and hence the number of augmentations
is often reduced.
The CAS algorithm performs scaling phases for successively smaller values
of a parameter~$\Delta$.
In a $\Delta$\dash scaling phase, each path augmentation delivers exactly $\Delta$
units of flow from a node~$v$ with at least $\Delta$ units of excess to a
node~$w$ with at least $\Delta$ units of deficit.
The shortest path searches are carried out in the so-called
\emph{$\Delta$\dash residual network}, which contains only those arcs whose
residual capacities are at least~$\Delta$.
When no such augmenting path is found, the value of $\Delta$ is halved
and the algorithm proceeds with the next phase.
Initially, $\Delta$ is set to $2^{\lfloor \log_2 U \rfloor}$ and the algorithm
terminates at the end of the phase in which $\Delta=1$.

The CAS algorithm maintains the reduced cost optimality conditions only in the
$\Delta$\dash residual network.
Each $\Delta$\dash scaling phase begins with saturating those newly introduced
arcs of the current $\Delta$\dash residual network that do not satisfy the
optimality conditions with respect to the current node potentials.
The saturations might increase the excess or deficit of some nodes,
but these requirements are satisfied in the subsequent phases.
At the end of the last phase, which corresponds to $\Delta = 1$, the
solution becomes both feasible and optimal since the $\Delta$\dash residual network
then coincides with the residual network.

In order to ensure the weakly polynomial running time of the CAS algorithm,
we need an additional assumption that a directed path of sufficiently large
capacity exists between each pair of nodes.
This condition, however, can easily be achieved by a simple extension of the
underlying network as follows.
Let $s$ denote a designated node of the network
(or a newly introduced artificial node).
For each other node~$i$, we can add new arcs $(i,s)$ and $(s,i)$ to the graph
with sufficiently large capacities and costs.
Under this additional assumption, the CAS algorithm is proved to solve the MCF
problem in $O(m\log U\cdot\textrm{SP}_+(n,m))$ time \cite{AMO93NetworkFlows}.

We made some modifications to this version of the CAS algorithm
in our implementation.
First, it is possible to avoid the above extension of the input graph
by allowing that more units of excess or deficit remain at the end of
a $\Delta$\dash scaling phase.
In this case, the polynomial running time bound is not proved, but our
experiments show that this version does not perform more path augmentations
and runs significantly faster in practice.
Therefore, our implementation does not extend the input graph by default.
Another modification utilizes that the path augmentations of each
$\Delta$\dash scaling phase might be capable of delivering more than
$\Delta$ units of flow.
We send the maximum possible amount of flow along each
path similarly to the SSP algorithm.
Furthermore, the scaling of the parameter~$\Delta$ can be
carried out using a factor other than two.
Let $\alpha\geq 2$ denote an integer scaling factor.
$\Delta$ is initially set to $\alpha^{\lfloor \log_\alpha U \rfloor}$
and divided by $\alpha$ at the end of each phase.
This means that more path augmentations might be required for the same
excess or deficit node in a $\Delta$\dash scaling phase, but the number of
phases is reduced.
In our experiments, a factor of $\alpha=4$ turned out to provide the
best overall performance, thus this option is used by default.

The CAS algorithm has much in common with the SSP method,
thus the practical improvements of the SSP implementation also applies to
this algorithm.
Our CAS code uses the same representations for the residual network and the
associated data.
In a $\Delta$\dash scaling phase, the $\Delta$\dash residual network is not
constructed explicitly, but the arcs with residual capacity less than
$\Delta$ are skipped during the path searches.
Moreover, our CAS implementation also terminates the shortest path computations
once an appropriate deficit node is permanently labeled and updates
the node potentials accordingly.
This idea and the practical data representations substantially improve the
performance of the CAS algorithm similarly to the SSP method.

\medskip
The computational results presented in Section~\ref{sec:testing} show
that the augmenting path algorithms, SSP and CAS, are not robust as
their performance greatly depends upon the characteristics of the input.
On general problem instances, these algorithms are typically slower than
the cost-scaling and network simplex methods, but
in certain cases, they turned out to be quite efficient.
For example, if the total excess is relatively small and hence
a few path augmentations are sufficient to solve the problem, the SSP algorithm
is usually the fastest method.

\subsection{Cost-scaling algorithm}
\label{sec:costscaling}
The cost-scaling technique for the MCF problem was proposed independently by
R\"ock \cite{Rock80Scaling} and
Bland and Jensen \cite{BlandJensen92Computational}.
Goldberg and Tarjan \cite{GoldbergTarjan90SuccApprox}
developed an improved method based on these algorithms by also utilizing the
concept of $\epsilon$\dash optimality,
which is due to Bertsekas \cite{Bertsekas79Distributed}
and, independently, Tardos~\cite{Tardos85StronglyPoly}.
The cost-scaling algorithm of Goldberg and Tarjan, which we henceforth
refer to as COS, can be viewed as a generalization of their well-known
push-relabel algorithm for the maximum flow problem
\cite{GoldbergTarjan88NewApproach}.
The COS algorithm is one of the most efficient solution methods
for the MCF problem, both in theory and practice.

The COS algorithm is a primal--dual method that applies a successive
approximation scheme by scaling upon the costs.
It iteratively produces $\epsilon$\dash optimal primal--dual
solution pairs for successively smaller values of $\epsilon\geq 0$.
(Recall the definitions and results related to $\epsilon$\dash optimality
from Section~\ref{sec:optimality}.)
Initially, $\epsilon=C$ and each phase preforms a \emph{refine}
procedure to transform an $\epsilon$\dash optimal solution into
an $(\epsilon/2)$\dash optimal solution until $\epsilon<1/n$.
At this stage, the algorithm terminates and Lemma~\ref{thm:epsopt_properties}
implies that an optimal flow is found.

The refine procedure takes an $\epsilon$\dash optimal primal--dual solution
pair $(x,\pi)$ as input and improves the approximation as follows.
First, it saturates each residual arc whose current reduced
cost is negative and thereby produces a pseudoflow~$x$ that is 0\dash optimal.
This means that $x$ is also $\epsilon$\dash optimal for any choice of $\epsilon$,
but it is not necessarily feasible.
After this step, the current approximation parameter $\epsilon$ is halved and
the pseudoflow~$x$ is gradually transformed into a feasible solution again,
but in a way
that preserves $\epsilon$\dash optimality for the new value of $\epsilon$.
This is achieved by performing a sequence of \emph{push} and \emph{relabel}
operations similarly to the push-relabel algorithm for the maximum flow problem.

Let $r_{ij}$ denote the residual capacity of an arc $(i,j)$
in the residual network~$G_x$ corresponding to the current pseudoflow~$x$
and let $e_i$ denote the signed excess value of node~$i$.
We call a node \emph{active} if its current excess is positive.
Furthermore, a residual arc $(i,j)$ is called \emph{admissible} if its
current reduced cost is negative and the subgraph of the residual network
consisting only of the admissible arcs is called the \emph{admissible network}.
The refine procedure throughout maintains
$\epsilon$\dash optimality and hence $-\epsilon\leq c_{ij}^\pi < 0$ holds for each
admissible arc $(i,j)$.
A~basic operation selects an active node~$i$
(i.e., $e_i>0$) and either \emph{pushes} flow on an admissible residual
arc $(i,j)$ or if no such arc exists, updates the potential of node~$i$,
which is called \emph{relabeling}.

A~push operation on an admissible residual arc $(i,j)$ is carried out by sending
$\delta = \min\{e_i, r_{ij}\}$ units of flow from node~$i$ to node~$j$
and thereby decreasing $e_i$ and increasing $e_j$ by~$\delta$.
This operation introduces the reverse arc $(j,i)$ into the residual network
unless it already had positive residual capacity, but this arc is not
admissible since $c_{ji}^\pi = -c_{ij}^\pi>0$.
If~an active node~$i$ has no admissible outgoing arc, a relabel operation
decreases its potential by~$\epsilon$.
This means that the reduced cost of each outgoing residual arc of node~$i$
is also decreased and the reduced cost of each incoming residual arc
is increased by~$\epsilon$.
Note that this modification preserves the $\epsilon$\dash optimality conditions
while creating new admissible outgoing arcs at node~$i$ and thus
allowing subsequent push operations to carry the excess of node~$i$.
Consequently, the only operation that can introduce a new admissible arc $(i,j)$
is the relabeling of node~$i$.
The refine procedure terminates when no active node remains in the network
and hence an $\epsilon$\dash optimal feasible solution is obtained.

It is proved that this generic version of the refine procedure performs
$O(n^2)$ relabel operations and $O(n^2m)$ push operations and hence runs in
$O(n^2m)$ time \cite{GoldbergTarjan90SuccApprox, AMO93NetworkFlows}.
Furthermore, the number of $\epsilon$\dash scaling phases is $O(\log(nC))$, as
$\epsilon$ is initially set to $C$ and it is halved at each phase until it
decreases below $1/n$.
Consequently, the generic COS algorithm runs in weakly polynomial time
$O(n^2m\log(nC))$.
Note, however, that the order in which the basic operations are performed
is not specified.
Goldberg and Tarjan \cite{GoldbergTarjan90SuccApprox} showed that
applying particular selection rules and using complex data structures
yield better theoretical running time.
They also developed a generalized framework to obtain a strongly polynomial
bound on the number of $\epsilon$\dash scaling phases by utilizing the same idea
that is exploited in the MMCC and CAT algorithms
(see Section~\ref{sec:cyclecanceling}).
The best variant they devised runs in $O(nm\log (n^2/m)\min\{\log(nC), m\log n\})$
time using dynamic trees
\cite{GoldbergTarjan90SuccApprox, SleatorTarjan83DynamicTree}.
Moreover, the COS algorithm turned out to be quite efficient in practice
and several complicated heuristics were also developed to improve its
performance~\cite{Goldberg97EfficientImpl}.

We implemented three variants of the COS method that perform the refine
procedure rather differently.

\paragraph{Push-relabel variant.}
This variant of the COS algorithm is based on the generic version
discussed above and hence performs local push and relabel operations in the
$\epsilon$\dash scaling phases.
We also applied several improvements and efficient heuristics in this
implementation according to the ideas found in
\cite{GoldbergTarjan90SuccApprox, AMO93NetworkFlows,
GoldbergKharitonov93Implementing,
Goldberg97EfficientImpl, BunnagelEtAl98EfficientImpl}.
In fact, most of these improvements and heuristics are analogous to
similar techniques devised for the push-relabel maximum flow algorithm.

The bottleneck of the COS algorithm corresponds to the searching of admissible
arcs for the basic operations.
Therefore, we applied the same graph representation that is used in
the augmenting path algorithms (see Section~\ref{sec:augpath}) as it is
intended to minimize the time required for iterating over the outgoing
residual arcs of a node.

Similarly to the capacity-scaling algorithm, the COS method also allows us
to use an arbitrary scaling factor $\alpha>1$.
We found that the optimal value depends on the problem, but
it was usually between 8 and 24 and the differences were moderate.
The default scaling factor is $\alpha=16$ in our implementation,
which typically performed very well.
Another practical modification targets the issue that the generic COS algorithm
performs internal computations with non-integer values of $\epsilon$ and
non-integer node potentials.
This drawback can be overcome by multiplying all arc costs by $\alpha n$
for a given integer scaling factor $\alpha\geq 2$ and by
scaling $\epsilon$ accordingly.
Initially, $\epsilon$ is set to $\alpha^{\lceil\log_\alpha(\alpha nC)\rceil}$
and is divided by $\alpha$ in each phase until $\epsilon = 1$.

We applied some improvements in the implementation of the refine
procedure, as well.
For performing the basic operations, we need to check the outgoing
residual arcs of each active node for admissibility.
These examinations can be made more efficiently if we record a
\emph{current arc} for each active node and continue the search for
an admissible outgoing arc from this current arc every time.
If an admissible arc is found, we perform a push operation and when we reach
the last outgoing arc of an active node without finding an
admissible arc, the node is relabeled and its current arc is set to
the first outgoing residual arc again.
(Recall that the definition of the basic operations imply that only the
relabeling of node~$i$ can introduce a new admissible arc outgoing from
node~$i$.)
Furthermore, the relabel operations are performed in a stricter way.
Instead of simply decreasing the potential of a relabeled node
by $\epsilon$, we decrease the potential by the largest possible amount that
does not violate the $\epsilon$\dash optimality conditions.
A single relabel operation thereby usually introduce more admissible arcs.
This modification significantly improves the overall performance of the algorithm
(up to a factor of two).

The strategy for selecting an active node for the next basic operation is
also important.
The number of active nodes is typically small, thus it is beneficial to
keep track of them explicitly.
A particular variant of the COS algorithm, known as the
\emph{wave implementation}, selects the active nodes according to a
topological ordering with respect to the admissible network.
This choice is proved to yield an $O(n^3)$\dash time implementation
of the refine procedure (instead of $O(n^2m)$).
However, our experiments showed that a simple FIFO selection rule using a
queue data structure usually results in less basic operations and
better performance in practice,
which is in accordance with \cite{BunnagelEtAl98EfficientImpl}
and~\cite{Goldberg97EfficientImpl}.

In addition to the implementation aspects discussed so far, some
effective heuristics can improve the practical performance of the COS
algorithm to a higher extent.
We implemented three such improvements out of the four
proposed by Goldberg \cite{Goldberg97EfficientImpl}.
These heuristics are also discussed in \cite{GoldbergKharitonov93Implementing}
along with detailed experimental evaluation.
Their practical effect depends on the problem instances
as well as the actual implementation and
the parameter settings (for example, the scaling factor~$\alpha$).

The \emph{potential refinement} (or \emph{price refinement}) heuristic
is based on the observation that an
$\epsilon$\dash scaling phase may produce a solution that is not only
$\epsilon$\dash optimal, but also $(\epsilon/\alpha)$\dash optimal
or even optimal.
Therefore, an additional step is introduced at the beginning of each phase
to check if the current solution is already $\epsilon$-optimal.
This heuristic attempts to adjust the potentials to satisfy the
$\epsilon$\dash optimality conditions, but without modifying the flow.
If $\epsilon$\dash optimality is verified, the refine procedure is skipped
and another phase is performed.
We implemented this potential refinement heuristic using an $O(nm)$\dash time scaling
shortest path algorithm \cite{Goldberg95ScalingAlgs} as suggested in
\cite{Goldberg97EfficientImpl}.
Our experiments also verified that this improvement usually eliminates the need
for the refine procedure in a few phases, especially the last ones.
Furthermore, the potential updates performed in this heuristic step typically
reduce the number of basic operations even in case the refine procedure
can not be skipped.
Consequently, this additional step significantly improves the overall
performance of the algorithm in most cases.

Another possible implementation of this heuristic performs a minimum-mean
cycle computation in each phase to determine the smallest $\epsilon$
for which the current flow is $\epsilon$-optimal and computes
corresponding node potentials.
This computation may allow us to skip more than one phase at once, but it is
usually slower, even using Howard's efficient algorithm.
Furthermore, this variant can be used to ensure a strongly polynomial bound
on the number of phases and thus on the overall running time, as well.
However, our experiments showed that the former variant of the potential refinement
heuristic, which we use in our final implementation,
is clearly superior to this one.
This result contradicts the conclusions of \cite{BunnagelEtAl98EfficientImpl}.

The \emph{global update} heuristic
performs relabel operations on several nodes in one step.
It iteratively applies the following \emph{set-relabel} operation.
Let $S\subset V$ denote a set of nodes such that it contains all deficit
nodes, but at least one active node is in $V\setminus S$.
If no admissible arc enters $S$, then the potential of every node in $S$
can be increased by $\epsilon$ without violating the
$\epsilon$\dash optimality conditions.
Furthermore, it is also shown in \cite{Goldberg97EfficientImpl}
that the theoretical running time of the COS algorithm remains unchanged
if the global update heuristic is applied only after every $\Omega(n)$
relabel operations.
In practice, this modification turned out to impose a huge improvement in
the efficiency of the algorithm on some problem classes, although
it does not help to much or even slightly worsens the performance
on other instances.
Our implementation of this heuristic follows the instructions presented in
\cite{BunnagelEtAl98EfficientImpl}.

The \emph{push-look-ahead} heuristic is another practical improvement for
the COS algorithm.
Its goal is to avoid pushing flow from node~$i$ to node~$j$ when a subsequent
push operation is likely to send this amount of flow back to node~$i$.
To achieve this, the maximum allowed amount of flow to be pushed into
a node~$i$ is limited by the sum of its deficit and the residual capacities
of its admissible outgoing arcs.
However, this idea requires the extension of the relabel operation to
those nodes at which this limitation is applied regardless of their current
excess values.
This heuristic is rather effective in practice, it usually decreases the
number of push operations significantly and hence the relabel operations
dominate the running time of the COS algorithm.
For more details about this heuristic,
see \cite{GoldbergKharitonov93Implementing,
Goldberg97EfficientImpl, BunnagelEtAl98EfficientImpl}.

Goldberg \cite{Goldberg97EfficientImpl} also suggests an additional
improvement, the \emph{arc fixing} heuristic.
The best version of this method speculatively fixes the flow values for
the arcs on which it is not likely to be changed later in the algorithm.
These arcs are excluded from the subsequent arc examinations, but
in certain cases, they have to be unfixed again.
We did not implement this heuristic yet, because it seems to be
rather involved and sensitive to parameter settings.
However, it would most likely improve the performance of our implementation.

We also remark that dynamic trees \cite{SleatorTarjan83DynamicTree}
can be used in the COS algorithm to perform a number of push operations at once,
which improves the theoretical running time \cite{GoldbergTarjan90SuccApprox}.
However, they are not likely to be practical due to the computational
overhead that these data structures usually impose and because
applying the above heuristics, the relabel operations become
the bottleneck of the algorithm instead of pushes
(see \cite{GoldbergKharitonov93Implementing, Goldberg97EfficientImpl}).
Therefore, we did not implement this variant.

\paragraph{Augment-relabel variant.}
This variant of the COS algorithm performs path augmentations instead of
local push operations, but relabeling
is heavily used to find augmenting paths.
At each step of the refine procedure, this method selects an active node~$v$
and performs a depth-first search in the admissible network to
find an augmenting path to a deficit node.
At an intermediate stage, the algorithm maintains an admissible path from an
active node~$v$ to the current node~$i$ and attempts to extend this path.
If node~$i$ has an admissible outgoing arc $(i,j)$, then the path is extended with
this arc and node~$j$ becomes the current node.
Otherwise, node~$i$ is relabeled and if $i\neq v$, we step back to the previous
node by removing the last arc of the current path.
When this search process reaches a deficit node, an augmenting path is found.

The flow augmentation on these admissible paths can be performed in
two different ways.
The first way is to push the same amount of flow on each arc of an
augmenting path, which is bounded by the smallest residual capacity
on the path as well as the excess of the starting node~$v$.
The other apparent implementation pushes the maximum possible amount of
flow on every arc of the path.
That is, for each arc $(i,j)$ of the path, $\delta_{ij}=\min\{e_i, r_{ij}\}$
units of flow is pushed on the arc, $e_i$ is decreased by $\delta_{ij}$,
and $e_j$ is increased by $\delta_{ij}$.
According to our experiments, this variant is slightly superior to the former one,
thus it is applied in our implementation.

Note that these path search and flow augmentation methods
correspond to a particular sequence of local push and relabel operations.
However, the actual push operations are carried out in a delayed and more guided
manner, in aware of an admissible path to a deficit node.
This helps to avoid such problems for which the push-look-ahead heuristic is
devised (see above), but a lot of work may be required to find augmenting paths,
especially if they are long.

Since this algorithm can be viewed as a special version of the generic COS
method, the same theoretical running time bound applies to it
as well as most of the practical improvements.
We used the same data representation, improvements and heuristics as for the
push-relabel algorithm except for the push-look-ahead heuristic,
which is obviously incompatible with this variant.
These modifications provided similar performance gains to those
measured for the push-relabel variant.

\paragraph{Partial augment-relabel variant.}
The third variant of the COS algorithm can be viewed as an intermediate
approach between the other two variants.
It~is based on the partial augment-relabel technique recently proposed by
Goldberg \cite{Goldberg08PartialAug} as an improvement for the push-relabel
maximum flow algorithm.
This method turned out to be more efficient and more robust than the
classical push-relabel algorithm and Goldberg also suggested the utilization of
the same idea in the MCF context.
According to the authors knowledge, our implementation of the COS algorithm
is the first to incorporate this technique.

The partial augment-relabel algorithm is quite similar to the augment-relabel
variant, but it limits the length of the augmenting paths.
The path search process is stopped either if a deficit node is reached
or if the length of the path reaches a given parameter $k\geq 1$.
In fact, the push-relabel and augment-relabel variants are
special cases of this approach for $k=1$ and $k=n$, respectively.
Goldberg \cite{Goldberg08PartialAug} suggests small values for the
parameter~$k$ in the maximum flow context,
which turned out to apply to the COS algorithm, as well.
In our experiments,
the optimal value of this parameter was typically between 3 and 8 and
the differences were not substantial for such small values of $k$.
Our default implementation uses $k=4$, just like Goldberg's maximum
flow implementation, as it turned out to be quite robust.

Apart from the length limitation for the augmenting paths, this
variant is exactly the same as the augment-relabel method.
(Actually, they have a common implementation but with different
values of the parameter $k$.)
However, the partial augment-relabel technique attains a good compromise between
the former two approaches and turned out to be clearly superior to them,
thus it is our default implementation of the COS algorithm.
Unless stated otherwise, we refer to this implementation as COS in the
followings.

\medskip
Section~\ref{sec:testing} provides experimental results for the COS
algorithm and its variants compared to other methods.
The classical push-relabel algorithm and especially the partial
augment-relabel variant using such heuristics and improvements
are highly efficient and robust in practice.
In contrast, the augment-relabel variant is often significantly slower.

\subsection{Network simplex algorithm}
\label{sec:networksimplex}
The primal \emph{network simplex} algorithm, which we henceforth refer
to as NS, is one of the most popular solution methods for
the MCF problem in practice.
It is a specialized version of the well-known linear programming (LP) simplex
method that exploits the network structure of the MCF problem and performs
the basic operations directly on the graph representation.
The LP variables correspond to the arcs of the graph and the LP bases
are represented by spanning trees.

The NS algorithm is devised by Dantzig, the inventor of the LP simplex method.
He first solved the uncapacitated transportation problem using this approach
and later generalized the bounded variable simplex method to directly solve
the MCF problem \cite{Dantzig63LP}.
Although the generic version of the NS algorithm does not run in polynomial
time, it turned out to be rather efficient in practice.
Therefore, subsequent research focused on efficient implementation
of the NS algorithm
\cite{BradleyEtAl77Design, BarrEtAl79TreeLabelling,
KenningtonHelgason80NetworkProg, Grigoriadis86EfficientImpl,
Lobel96NetworkSimplex}
as well as on developing special variants of both the primal and the dual
network simplex methods that run in polynomial time
\cite{Tarjan91Efficiency, GoldfarbHao92polynomial,
OrlinEtAl93PolyDual, Orlin97PolyPrimal, Tarjan97DynamicTreesNetwork}.
Detailed discussion of the NS method considering both
theoretical and practical aspects can be found in
\cite{AMO93NetworkFlows} and \cite{KellyOneill91NetworkSimplex}.

The fundamental concept on which the NS algorithm is based
is the notion of \emph{spanning tree solutions}.
Such a solution is represented by a partitioning of the node set~$V$
into three subsets $(T,L,U)$ such that
each arc in $L$ has flow fixed at zero (\emph{lower bound}),
each arc in $U$ has flow fixed at the capacity of the arc
(\emph{upper bound}),
and the arcs in $T$ form an (undirected) spanning tree of the network.
The flow on these \emph{tree arcs} also satisfy the nonnegativity and
capacity constraints, but they are not restricted to any of the bounds.
It can easily be seen that the flow values on the tree arcs are uniquely
determined by the partitioning $(T,L,U)$ since there is no cycle in $T$.
Furthermore, it is proved that if an instance of the MCF problem has an optimal
solution, then it also has an optimal spanning tree solution, which
can be found by successively transforming a spanning tree solution to another
(see \cite{AMO93NetworkFlows}).
Actually, these spanning tree solutions correspond to the LP basic feasible
solutions of the problem.
This observation allows us to implement the simplex method by performing
all operations directly on the network, without maintaining the simplex tableau,
which makes this approach very efficient.

The standard simplex method maintains a basic feasible solution and
gradually improves its objective function value by small transformations,
known as \emph{pivots}.
Accordingly, the NS algorithm throughout maintains a spanning tree solution
of the MCF problem and successively decreases the total cost of the flow
until it becomes optimal.
Furthermore, node potentials are also maintained such that the reduced cost
of each arc in the spanning tree equals to zero.
At each step, a non-tree arc violating its complementary slackness optimality
condition (see Theorem~\ref{thm:compslack_optimality}) is added to
the current spanning tree, which uniquely determines a negative cost
residual cycle.
This cycle is then canceled by augmenting the maximum possible amount of
flow on it and a tree arc corresponding to a saturated residual arc is
selected to be removed from the tree.
The node potentials are also adjusted to preserve the property that
the reduced costs of each tree arc is zero and
finally, the tree structure is updated.
This whole operation transforming a spanning tree solution to another
is called \emph{pivot}.
If no suitable entering arc can be found, the current flow is optimal
and the algorithm terminates.

In fact, the NS algorithm can also be viewed as a particular variant of the
cycle-canceling method (see Section~\ref{sec:cyclecanceling}).
Due to the sophisticated method of maintaining spanning tree solutions, however,
a negative cycle can be found and canceled much faster (in linear time).
On the other hand, an additional technical issue, known as \emph{degeneracy},
may arise in the NS algorithm.
If the spanning tree contains an arc whose flow value equals to zero or the
capacity of the arc, then a pivot step may detect a cycle of zero
residual capacity.
Such \emph{degenerate} pivots only modify the spanning tree, but the flow
itself remains unchanged.
Consequently, it is possible that several consecutive pivots do not actually
decrease the flow cost (known as \emph{stalling})
or, which is even worse, the same
spanning tree solution occurs multiple times and hence the algorithm
does not necessarily terminate in a finite number of iterations
(known as \emph{cycling}).
Experiments with certain classes of large-scale MCF problems showed that
more than 90\% of the pivots may be degenerate.

A simple and popular technique to overcome such difficulties
is based on the concept of
\emph{strongly feasible spanning tree solutions}.
A spanning tree solution is called strongly feasible if a positive amount of
flow can be sent from each node to a designated root node of the
spanning tree along the tree path without violating the nonnegativity and
capacity constraints.
Using an appropriate rule for selecting the leaving arcs, the NS
algorithm can throughout maintain a strongly feasible spanning tree.
This technique is proved to ensure that the algorithm terminates
in a finite number of iterations \cite{AMO93NetworkFlows}.
Furthermore, it substantially decreases the number of degenerate pivots
in practice and hence makes the algorithm faster.

It can be shown, using a perturbation technique,
that the NS algorithm maintaining a strongly feasible spanning tree solution
performs $O(nmCU)$ pivots for the MCF problem with integer data
regardless of the selection rule of entering arcs
\cite{AMO89NetworkFlows}.
An entering arc can be found in $O(m)$ time and using an appropriate
labeling technique, the spanning tree structure can be updated in $O(n)$
time.
Therefore, a single pivot takes $O(m)$ time and the total running time
of the NS algorithm is $O(nm^2CU)$.
However, this bound does not reflect to the typical performance of the
algorithm in practice.

The implementation of the primal NS algorithm is based on a practical
storage scheme of the spanning tree solutions that makes it possible
to perform the basic operations of the algorithm efficiently.
The book of Kennington and Helgason \cite{KenningtonHelgason80NetworkProg}
discusses several such spanning tree data structures along with methods for
updating them during the iterations of the algorithm.
A quite popular approach, sometimes referred to as the ATI
(\emph{Augmented Threaded Index}) method, represents a spanning tree as follows.
The tree has a designated root node and three indices is stored for each
node in the tree:
the \emph{depth} of the node (i.e., the distance from the root node in the tree),
the \emph{parent} of the node in the tree, and
a \emph{thread} index which is used to define a depth-first traversal of the
spanning tree.
This storage scheme and its update mechanism are discussed in detail in
\cite{KellyOneill91NetworkSimplex} and \cite{AMO93NetworkFlows}.
The ATI technique has an improved version, which is due to
Barr, Glover, and Klingman~\cite{BarrEtAl79TreeLabelling} and is
usually referred to as the XTI (\emph{eXtended Threaded Index}) method.
The XTI scheme replaces the \emph{depth} index by two indices for each node:
the \emph{number of successors} of the node in the tree and
the \emph{last successor} of the node according to the traversal
defined by the \emph{thread} index.
Other approaches, for example the so-called API and XPI methods, are
also often applied.

We implemented the primal NS algorithm using both ATI and XTI techniques.
The latter one has two advantages over the ATI method.
First, the XTI indices can be updated more efficiently since a tree alteration
of a single pivot usually modifies the depth of several nodes in subtrees that
are moved from a position to another, while the set of successors is
typically modified only for a much smaller number of nodes.
In addition, the XTI method also allow an improved updating process for
the node potentials.
Note that by removing the leaving arc, the current spanning tree is
divided into two subtrees, which are then connected again by the entering arc.
In order to preserve zero reduced costs for the tree arcs, we have to
increase the potential of each node in one of the subtrees
by a certain constant value~$\lambda$ or decrease the potential of
each node in the other subtree by $\lambda$.
The XTI scheme makes it possible to immediately determine which subtree is
smaller and to perform the update process on the smaller subtree.

Although the XTI labeling method is not as widely known and popular as the simpler
ATI method, our experiments showed that it is much more efficient than
ATI on all problem instances.
Therefore, the final version of our code only implements the XTI technique.
The substantial performance gain of this approach is due to the
first advantage mentioned above.
In contrast, we found that the alternative potential update is not so
important, because the subtree containing the root node turned out to be
the bigger one in virtually all pivots.
Moreover, we can easily avoid overflow problems related to node potentials
and reduced costs if the potential of the root node is not modified
throughout the algorithm.
Therefore, we decided to update potentials in the subtree not containing
the root node in every step.
We also applied an important improvement in the implementation of
the XTI method.
An additional \emph{reverse thread} index is also stored for each node
and hence the depth-first traversal is represented by a doubly-linked list.
This modification turned out to substantially improve the performance of
the update process.
In fact, the inventors of the XTI technique also discussed this improvement
\cite{BarrEtAl79TreeLabelling}, but they did not applied it to reduce
the memory requirements of the representation.
(However, note that enormous progress has been made on the computers
since the time when that paper was written.)

Another interesting aspect of the data representation for the NS algorithm is
that we need not traverse the incident arcs of nodes throughout the algorithm,
although such examinations are crucial in other algorithms.
Therefore, we applied a quite simple and unusual graph representation to
implement the NS algorithm.
The nodes and arcs are represented by consecutive integers and we store the
source and target nodes for each arc (in arrays), but we do not keep track
of the incident arcs of a node at all.

The NS algorithm also requires an initial spanning tree solution to start with.
It is possible to transform any feasible solution~$x$ to a spanning tree
solution~$x'$ such that the total cost of $x'$ is less than or equal to
the total cost of $x$.
Furthermore, the required spanning tree indices can also be computed by
a depth-first traversal of the tree arcs.
However, artificial initialization techniques are much more common in practice.
An artificial root node~$s$ with zero supply is typically added to the network
as well as artificial arcs with sufficiently large capacities and costs.
Recall that the signed supply value of node~$i$ is denoted as $b_i$.
For each original node~$i$, we add a new arc $(i,s)$
to the network if $b_i\geq 0$ and add a new arc $(s,i)$ otherwise.
We can set the capacity of each new arc to $nU$ and its cost to $nC$.
In this extended network, we can easily construct a strongly feasible
spanning tree solution~$x$ as follows.
For each original arc $(i,j)$, let $x_{ij}=0$ and for each original
node~$i$, let $x_{is}=b_i$ if $b_i\geq 0$ and let $x_{si}=-b_i$ otherwise.
The initialization of the tree indices and the node potentials is
straightforward in this case.
Furthermore, note that an optimal solution in the extended network does not
send flow on artificial arcs due to their large costs
unless the original problem is infeasible.

We experimented with both ways of initialization and it turned out that
the artificial method usually provides better overall performance
mainly because of two reasons.
First, the artificial spanning tree solution can be constructed easily and quickly.
Second, it allows efficient tree update for the first few pivots
as the depth of the tree is rather small.
Therefore, we decided to use only this variant in our final implementation.
The strongly feasibility is preserved throughout the algorithm by
carefully selecting the leaving arc whenever multiple residual arcs
are saturated by a pivot step
(see \cite{KellyOneill91NetworkSimplex, AMO93NetworkFlows} for details).

One of the most crucial aspects of the NS algorithm,
which is not considered so far, is the selection the entering arcs.
Recall that the reduced cost of each tree arc is zero and each non-tree arc has
a flow value fixed either at zero or the capacity of the arc.
Therefore, a non-tree arc $(i,j)$ allows flow augmentation only in one direction.
If the reduced cost of the residual arc associated with this direction is negative,
the arc $(i,j)$ can be selected to enter the tree.
In this case, a negative-cost residual cycle is formed by this arc and
the unique tree path connecting nodes $i$ and $j$
(these tree arcs have zero reduced costs).
To~implement the NS algorithm, we require a method for selecting such an
entering arc at each iteration, which is usually referred to
as \emph{pivot rule} or \emph{pricing strategy}.
The applied method
affects the ``goodness'' of the entering arcs and thereby the number of
iterations as well as the average time required for selecting an entering arc,
which is a dominant part of each iteration.
Consequently, applying different strategies, we can obtain several variants of the
NS algorithm with quite different theoretical and empirical behavior.

We implemented five pivot rules, which are discussed in the followings.
Four of them are widely known and well-studied rules
\cite{KellyOneill91NetworkSimplex, AMO93NetworkFlows}, while the fifth
one is an improved version of the \emph{candidate list} rule.
In the discussion of these methods, a non-tree arc is called \emph{eligible}
if it does not satisfy the complementary slackness optimality condition
and hence can be selected as an entering arc.
Let $\pi$ denote the current set of node potentials and let $c_{ij}^\pi$
denote the reduced cost of an arc $(i,j)$.
An eligible arc $(i,j)$ either has zero flow and $c_{ij}^\pi<0$ or
has a flow equal to its capacity and $c_{ij}^\pi>0$.
We refer to $|\,c_{ij}^\pi\,|$ as the \emph{violation} of an eligible arc $(i,j)$.

\paragraph{Best eligible arc pivot rule.}
This is one of the simplest and earliest pivot strategies, which was proposed
by Dantzig and is also known as \emph{Dantzig's pivot rule}.
At each iteration, this method selects an eligible arc with the maximum
violation to enter the tree.
This means that a residual cycle having the most negative total cost is
selected to be canceled, which causes the maximum decrease of the
objective function per unit flow augmentation.
Computational studies showed that this selection rule usually
results in fewer iterations than other strategies.
However, it has to consider all non-tree arcs and recompute their
reduced costs to select the best eligible arc at each iteration.
Consequently, the overall performance of the NS algorithm with this pivot
rule is rather poor despite the small number of iterations.

\paragraph{First eligible arc pivot rule.}
Another straightforward idea is to select the first eligible arc at
each iteration.
The practical implementation of this rule examines the arcs cyclically by
starting each search process at the position where the previous eligible
arc is found.
If we reach the end of the arc list, the examination is continued
from the beginning of the list again.
If a pivot operation examines all non-tree arcs without finding an
eligible arc, the solution is optimal and the algorithm terminates.
This strategy represents the other extreme way of selecting the entering arcs
compared to the previous rule.
It rapidly finds an entering arc at each iteration, but these arcs
typically have relatively small violation and hence a lot of iterations
are usually required.

\paragraph{Block search pivot rule.}
Since the previous two rules do not perform well in practice, several other
strategies have been devised to implement effective compromise between them.
A simple \emph{block search} approach is proposed by
Grigoriadis \cite{Grigoriadis86EfficientImpl}.
This method cyclically examines blocks of arcs and
selects the best eligible candidate among these arcs at each iteration.
The search process starts from the position of the previous entering arc
and checks a specified number of arcs by recomputing their reduced costs.
If this block contains eligible arcs, then the one with the maximum
violation is selected to enter the basis.
Otherwise, we examine one or more subsequent blocks of arcs
until an eligible arc is found.

The block size~$B$ is an important parameter of this method.
In fact, the previous two rules are special cases of this one with
$B=m$ and $B=1$.
Several sources suggest to set $B$ proportionally to the number of arcs,
for example, between 1\% and 10\%
\cite{Grigoriadis86EfficientImpl, KellyOneill91NetworkSimplex}.
However, our experiments clearly showed that much better overall performance
can be achieved on virtually all problem classes if we set
$B=\alpha\sqrt{m}$ for small values of $\alpha$
(for example, between 0.5 and~2).
In our implementation, $B=\sqrt{m}$ is used, which results in a highly
efficient and robust pivot rule.

Similarly to the \emph{first eligible rule}, this strategy also has the
inherent advantage that an arc is allowed to enter the basis
only periodically, which usually decreases the number of degenerate
pivots in practice
\cite{Cunningham79Theoretical, GoldfarbEtAl90Antistalling}.

\paragraph{Candidate list pivot rule.}
This is another classical pivot rule, which was proposed by
Mulvey \cite{Mulvey78Pivot}.
It occasionally builds a list of eligible arcs and selects the best
arcs among these candidates at subsequent iterations.
A so-called \emph{major} iteration examines the arcs in a wraparound fashion
similarly to the previous rules and builds a list containing at most $L$
eligible arcs.
After a major iteration, we perform at most $K$ \emph{minor} iterations,
each of which scans this list and selects an eligible arc with maximum
violation to enter the basis.
If an arc is not eligible any more, it is removed from the list.
When $K$ minor iterations are performed or the list becomes empty,
another major iteration takes place.

This method is similar to the \emph{block search rule}, but it considers
the same subset of the arcs in several consecutive pivots,
while the previous rule considers only the best arc of a block and then
advances to the next block.
We obtained the best average running time for this rule
using $L=\sqrt{m}/4$ and $K=L/10$.
However, this method usually performed worse than the simpler
block search strategy.

\paragraph{Altering candidate list pivot rule.}
This strategy was developed by us as an improved version of the
\emph{candidate list rule}.
There are various other rules that exploit similar ideas, but the authors
are not aware of another implementation of this method.
It maintains a candidate list similarly to the previous rule, but it
attempts to extend this list at each iteration and keeps only the several best
candidates of the previous iterations.
The candidate arcs are collected with the search process used in the
\emph{block search rule}.

This method has two parameters: a block size $B$ and the maximum length of
the altering candidate list, which is denoted by $H$.
At the beginning of each iteration, we check the current candidate list
and remove all arcs that are not eligible any more.
After that, at least one arc block of size $B$ is examined to extend the candidate
list with new eligible arcs.
If a nonempty list is obtained, then an arc of maximum violation is selected
from the list to enter that basis.
The other arcs are then partially sorted and the list is truncated to
contain at most $H$ of the best candidates in terms of their current violation.
According to our experiments, this method is very efficient using
$B=\sqrt{m}$ and $H=B/100$.

\medskip
Numerous other rules have also been developed applying similar or more
complicated partial pricing techniques.
We also implemented several variants, but the \emph{block search pivot rule}
and the \emph{altering candidate list pivot rule} turned out to provide the
best overall performance.
Since the block search rule is simpler and turned out to be slightly
more robust, it is our default pivot strategy.
Unless stated otherwise, we refer to this variant as NS in the followings.

We also developed an additional heuristic based on the artificial
initialization procedure of the algorithm to make the first few pivots faster.
The initialization of node potentials implies that an arc $(i,j)$
is eligible for the first pivot if and only if $b_i\geq 0$ and $b_j<0$.
After such an arc enters the basis, new arcs incident to its
source node may also become eligible.
Therefore, we collect several arcs using a partial traversal of the graph
starting from the demand nodes and using the reverse orientation of each arc.
This arc list is then used by the first few pivots to select entering arcs from.
Our computational results showed that this idea slightly improves the overall
performance of the NS algorithm by making these pivots substantially faster.

Section~\ref{sec:testing} provides experimental results comparing the
different pivot rules as well as comparing the NS algorithm to other
solution methods.
Our NS implementation turned out to be highly efficient, especially on
relatively small and medium-sized networks, but it is typically outperformed by
the cost-scaling codes on the largest problem instances.

\section{Experimental study}
\label{sec:testing}
This section presents an empirical study of the implemented algorithms
and also compares them to other efficient MCF solvers.
The contribution of these results is twofold.
First, a great number of MCF algorithms are compared using the same benchmark
suite, which provides insight into their relative performance on different
classes of networks.
Second, larger problem instances are also considered than that in previous
experimental studies of MCF algorithms, which turned out to be important
to draw reliable conclusions related to the asymptotic behavior of
these algorithms.

The experiments were conducted on a machine with
AMD \mbox{Opteron} \mbox{Dual} \mbox{Core} 2.2~GHz CPU and
16~GB RAM (1~MB cache), running \mbox{openSUSE 11.4} operating system.
All codes were compiled with GCC~4.5.3 using \mbox{-O3} optimization flag.

\subsection{Test instances}
\label{sec:instances}
Our test suite contains numerous problem instances of
variable size and characteristics.
Most of these instances were generated with standard random generators,
NETGEN and GOTO, while the others are based on either real-life road networks
or maximum flow problems arising in computer vision applications.
The largest networks contain millions of nodes and arcs.

\paragraph{NETGEN instances.}
NETGEN \cite{KlingmanEtAl74Netgen} is a classical generator that produces random
instances of the MCF problem and other network optimization problems.
It is generally known to produce relatively easy MCF instances.
The source code of NETGEN is available at the FTP site of the First DIMACS
Implementation Challenge \cite{DIMACS1FTP}.

Our benchmark suite contains four problem families created with NETGEN.

\begin{itemize}
  \item\textbf{NETGEN-8.}
  This family contains sparse networks, for which the average outdegree
  of the nodes is 8 (i.e., $m=8n$).
  The arc capacities and costs are selected uniformly at random from the ranges
  [1..1000] and [1..10000], respectively.
  The number of supply and demand nodes are both set to about $\sqrt{n}$
  and the average supply per supply node is 1000.

  \item\textbf{NETGEN-SR.}
  This family contains relatively dense networks, for which the average
  outdegree is about $\sqrt{n}$ (i.e., $m\approx n\sqrt{n}$).
  The other parameters are set the same way as for the NETGEN-8 family.

  \item\textbf{NETGEN-LO-8.}
  This family is similar to NETGEN-8 with the only difference that the
  average supply per supply node is much lower, namely 10 instead of 1000.
  Therefore, the arc capacities incorporate only ``loose'' bounds for the
  feasible solutions.

  \item\textbf{NETGEN-DEG.}
  In these instances, the number of nodes is fixed to $n=4096$ and the average
  outdegree ranges from $2$ to $n/2$.
  Other parameters are the same as of NETGEN-8 and NETGEN-SR instances.
\end{itemize}

\paragraph{GOTO instances.}
GOTO is another well-known random generator for the MCF problem, which is
intended to produce hard instances.
It is developed by Goldberg and is described in \cite{GoldbergKharitonov93Implementing}.
The name of the generator stands for \emph{Grid On Torus}, which reflects to the
basic structure of the generated networks.
Each GOTO problem instance has one supply node and one demand node and the
supply value is adjusted according to the arc capacities.
This generator is also available at \cite{DIMACS1FTP}.

We used two GOTO families in our experiments, which differ only in the density of
the networks.

\begin{itemize}
  \item\textbf{GOTO-8.}
  This family consists of sparse networks with an average outdegree of 8.
  Similarly to the NETGEN families,
  the maximum arc capacity is set to 1000, while the maximum arc cost is
  set to 10000.

  \item\textbf{GOTO-SR.}
  This family consists of relatively dense networks with an average outdegree of
  about $\sqrt{n}$. Other parameters are the same as of the GOTO-8 family.
\end{itemize}

\paragraph{ROAD instances.}
We also experimented with MCF problems that are based on real-world road networks.
To generate such instances, we used the TIGER/Line road network files of
several states of the USA.
These data files are available at the web site of the Ninth DIMACS
Implementation Challenge~\cite{DIMACS9WWW}.

In our experiments, we selected seven states with road networks of increasing size,
namely DC, DE, NH, NV, WI, FL, and TX, and generated MCF problem instances
as follows.
The original undirected graphs are converted to directed graphs by replacing
each edge with two oppositely directed arcs.
The cost of an arc is set to the travel time on the corresponding road section and
the arc capacities are uniformly set to one.
This means that we are actually looking for a specified number of arc-disjoint
directed paths from supply nodes to demand nodes having minimum total cost.
The number of supply and demand nodes are both $\lfloor\sqrt{n}/10\rfloor$.
These nodes are selected randomly and the supply-demand values are determined
by a maximum flow computation that maximizes the total supply with respect to
the fixed set of supply and demand nodes.

\paragraph{VISION instances.}
This family consists of MCF instances based on large-scale
maximum flow problems arising in computer vision applications.
These maximum flow data files were made available
at \url{http://vision.csd.uwo.ca/data/maxflow/} by the
Computer Vision Research Group at the University of Western Ontario.
They are intended to be used for benchmarking maximum flow algorithms
(for example, see \cite{Goldberg08PartialAug}).

We used some of the segmentation instances related to
medical image analysis.
These instances are defined on three-dimensional grid networks.
We selected those variants in which the underlying networks are 6\dash connected and
the maximum arc capacity is 100 (namely, the \texttt{bone\_sub*\_n6c100} files).
These maximum flow instances were converted to minimum-cost maximum flow
problems using random arc costs selected uniformly from the range [1..100].
The original networks also contain arcs of zero capacity,
but we skipped these arcs during the transformation and hence
did not preserve the 6\dash connectivity.

\medskip
We also experimented with several other problem instances and generator parameters,
but this collection turned out to be a representative benchmark suite of
reasonable size.
For all problem families, we generated three instances of each problem size
with different random seeds.
In all cases, we report the average running time over such three instances
to provide more relevant results.

\subsection{Comparison of the implemented algorithms}
\label{sec:comparison_lemon}
This subsection presents benchmark results for the implemented algorithms
and their variants.
Each table reports running time results in seconds.
The size of a problem instance is indicated by the number of
nodes~$n$ and the average outdegree $deg$ (i.e., $m = deg\cdot n$).
The best running time is highlighted for each problem size.
The codes were executed with an explicit timeout limit of one hour
and a ``$-$'' sign denotes the cases when this timeout limit was reached.
Some charts are also presented showing running time as a
function of the number of nodes in the network
(logarithmic scale is used for both axes).

\begin{figure}[!htb]
  \begin{minipage}{.5\textwidth}
    \centerline{\footnotesize\hspace{1em} NETGEN-8 family ($m=8n$)}\vspace{.5ex}
    \includegraphics[width=0.98\textwidth]{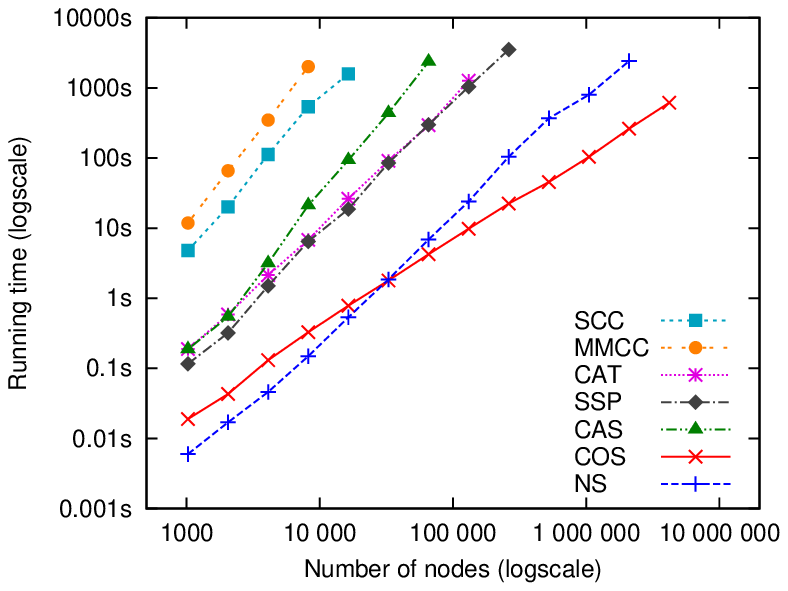}
  \end{minipage}
  \begin{minipage}{.5\textwidth}
    \centerline{\footnotesize\hspace{1em} NETGEN-SR family ($m\approx n\sqrt{n}$)}\vspace{.5ex}
    \includegraphics[width=0.98\textwidth]{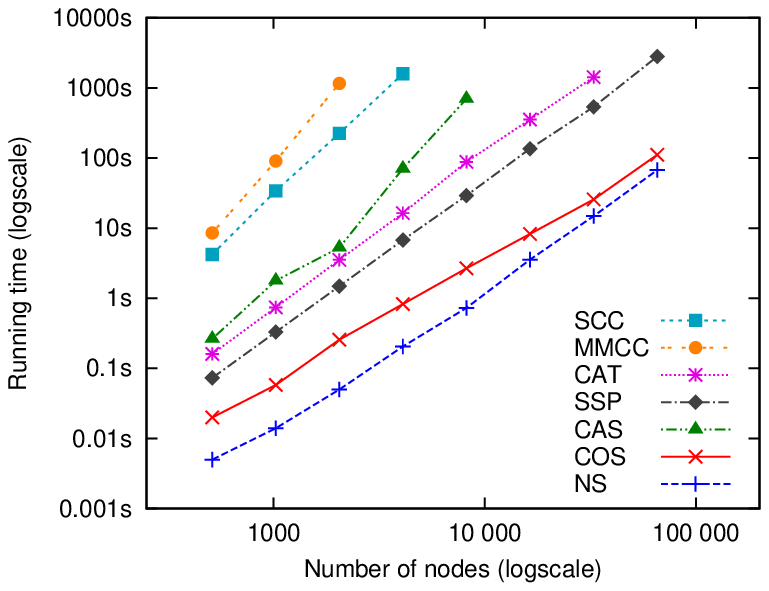}
  \end{minipage}
  \vspace{-1.5ex}
  \caption{Comparison of our implementations on NETGEN instances}
  \label{fig:lemon_algs_netgen}
\end{figure}

Figure~\ref{fig:lemon_algs_netgen} and
Tables~\ref{tab:lemon_algs_netgen_8} and \ref{tab:lemon_algs_netgen_sr}
present the running time results in seconds for NETGEN\dash 8 and
NETGEN\dash SR families.
On these instances, the SCC algorithm was about 3 times faster than MMCC,
while the CAT algorithm greatly outperformed both of them.
It is quite interesting that the simple SSP algorithm was
an order of magnitude faster than its capacity-scaling variant, CAS.
The CAT algorithm performed similarly to SSP on NETGEN\dash 8
instances, but was significantly slower on NETGEN\dash SR networks.
The COS and NS algorithms were the most efficient on these instances.
They turned out to be orders of magnitude faster than the other algorithms.
COS showed better asymptotic behavior than NS
(see Figure~\ref{fig:lemon_algs_netgen}) and was significantly
faster on the largest NETGEN\dash 8 instances.
On the other hand, NS was more efficient on the relatively
small sparse networks and on all NETGEN\dash SR instances
despite its worse asymptotic trends.

\begin{table}[!htb]
  \centering
  \footnotesize
  \begin{tabular}{ccrrrrrrr}
   \hline \\[-2.25ex]
            $n$ & $deg$ &         SCC &        MMCC &         CAT &         SSP &         CAS &         COS &         NS \\
   \hline \\[-2.25ex]
    \pow{2}{10} &     8 &       4.81  &      11.85  &       0.19  &       0.12  &       0.19  &       0.02  &    \x{0.01} \\
    \pow{2}{12} &     8 &     112.40  &     347.47  &       2.14  &       1.50  &       3.15  &       0.13  &    \x{0.05} \\
    \pow{2}{14} &     8 &    1587.01  &        $-$  &      26.36  &      18.66  &      93.14  &       0.78  &    \x{0.54} \\
    \pow{2}{16} &     8 &        $-$  &        $-$  &     295.05  &     298.95  &    2360.21  &    \x{4.24} &       6.88  \\
    \pow{2}{18} &     8 &        $-$  &        $-$  &        $-$  &    3514.72  &        $-$  &   \x{22.40} &     104.69  \\
    \pow{2}{20} &     8 &        $-$  &        $-$  &        $-$  &        $-$  &        $-$  &  \x{103.83} &     799.26  \\
    \pow{2}{22} &     8 &        $-$  &        $-$  &        $-$  &        $-$  &        $-$  &  \x{615.42} &        $-$  \\
   \hline
  \end{tabular}
  \caption{Comparison of our implementations on NETGEN-8 instances}
  \label{tab:lemon_algs_netgen_8}
\end{table}

\begin{table}[!htb]
  \centering
  \footnotesize
  \begin{tabular}{crrrrrrrr}
   \hline \\[-2.25ex]
            $n$ & $deg$ &         SCC &        MMCC &         CAT &         SSP &         CAS &         COS &         NS \\
   \hline \\[-2.25ex]
    \pow{2}{10} &    32 &      34.03  &      90.82  &       0.74  &       0.33  &       1.79  &       0.06  &    \x{0.01} \\
    \pow{2}{11} &    45 &     224.79  &    1158.92  &       3.54  &       1.49  &       5.27  &       0.26  &    \x{0.05} \\
    \pow{2}{12} &    64 &    1592.62  &        $-$  &      16.36  &       6.77  &      70.35  &       0.83  &    \x{0.21} \\
    \pow{2}{13} &    91 &        $-$  &        $-$  &      88.13  &      29.16  &     697.47  &       2.68  &    \x{0.73} \\
    \pow{2}{14} &   128 &        $-$  &        $-$  &     353.23  &     136.17  &        $-$  &       8.28  &    \x{3.55} \\
    \pow{2}{15} &   181 &        $-$  &        $-$  &    1419.60  &     535.57  &        $-$  &      25.74  &   \x{14.90} \\
    \pow{2}{16} &   256 &        $-$  &        $-$  &        $-$  &    2799.34  &        $-$  &     111.55  &   \x{67.29} \\
   \hline
  \end{tabular}
  \caption{Comparison of our implementations on NETGEN-SR instances}
  \label{tab:lemon_algs_netgen_sr}
\end{table}

\newpage

Table~\ref{tab:lemon_algs_netgen_lo_8} contains performance results for the
NETGEN\dash LO\dash 8 family.
As~one would expect, these instances turned out to be easier to solve
than NETGEN\dash 8 instances of the same size.
For this family, SSP was an order of magnitude faster than CAT, while
CAS was even more efficient than the SSP algorithm by a factor between 2 and 3.
Note that the relative performance of SSP and CAS is entirely different
compared to the NETGEN\dash 8 results.
Nevertheless, the fastest methods were COS and NS just like for the NETGEN\dash 8
family and their relationship was similar.

\begin{table}[!htb]
  \centering
  \footnotesize
  \begin{tabular}{ccrrrrrrr}
   \hline \\[-2.25ex]
            $n$ & $deg$ &         SCC &        MMCC &         CAT &         SSP &         CAS &         COS &         NS \\
   \hline \\[-2.25ex]
    \pow{2}{10} &     8 &       0.82  &       2.52  &       0.14  &       0.02  &       0.01  &       0.01  &    \x{0.00} \\
    \pow{2}{12} &     8 &       8.49  &      79.93  &       1.88  &       0.13  &       0.06  &       0.07  &    \x{0.02} \\
    \pow{2}{14} &     8 &      73.83  &    1801.08  &      22.10  &       1.29  &       0.67  &       0.43  &    \x{0.19} \\
    \pow{2}{16} &     8 &     668.00  &        $-$  &     183.06  &      17.79  &       6.67  &       2.65  &    \x{2.11} \\
    \pow{2}{18} &     8 &        $-$  &        $-$  &    2062.12  &     172.63  &      60.16  &   \x{13.79} &      29.00  \\
    \pow{2}{20} &     8 &        $-$  &        $-$  &        $-$  &    1342.73  &     519.06  &   \x{68.41} &     293.49  \\
    \pow{2}{22} &     8 &        $-$  &        $-$  &        $-$  &        $-$  &        $-$  &  \x{457.02} &    2482.50  \\
   \hline
  \end{tabular}
  \caption{Comparison of our implementations on NETGEN-LO-8 instances}
  \label{tab:lemon_algs_netgen_lo_8}
\end{table}

Table~\ref{tab:lemon_algs_netgen_deg} shows how the running time of the algorithms
depends on the density of the network.
NS was clearly the most efficient algorithm in these tests.
COS and SSP were also relatively fast, while CAT turned out to be significantly slower
and the other methods were not competitive.
The CAS algorithm performed much worse than SSP on the dense instances.

\begin{table}[!htb]
  \centering
  \footnotesize
  \begin{tabular}{crrrrrrrr}
   \hline \\[-2.25ex]
            $n$ & $deg$ &         SCC &        MMCC &         CAT &         SSP &         CAS &         COS &         NS \\
   \hline \\[-2.25ex]
    \pow{2}{12} &     2 &      18.43  &      60.71  &       1.03  &       0.33  &       0.27  &       0.08  &    \x{0.02} \\
    \pow{2}{12} &     8 &     112.40  &     347.47  &       2.14  &       1.50  &       3.15  &       0.13  &    \x{0.05} \\
    \pow{2}{12} &    32 &     657.29  &    2655.80  &       7.41  &       4.40  &      35.54  &       0.43  &    \x{0.13} \\
    \pow{2}{12} &   128 &    3523.55  &        $-$  &      32.03  &      12.73  &     273.76  &       1.98  &    \x{0.39} \\
    \pow{2}{12} &   512 &        $-$  &        $-$  &     147.58  &      35.58  &    2473.16  &       7.72  &    \x{1.31} \\
    \pow{2}{12} &  2048 &        $-$  &        $-$  &     624.44  &      99.49  &        $-$  &      47.08  &    \x{6.68} \\
   \hline
  \end{tabular}
  \caption{Comparison of our implementations on NETGEN-DEG instances}
  \label{tab:lemon_algs_netgen_deg}
\end{table}

\begin{figure}[!htb]
  \begin{minipage}{.5\textwidth}
    \centerline{\footnotesize\hspace{1em} GOTO-8 family ($m=8n$)}\vspace{.5ex}
    \includegraphics[width=0.98\textwidth]{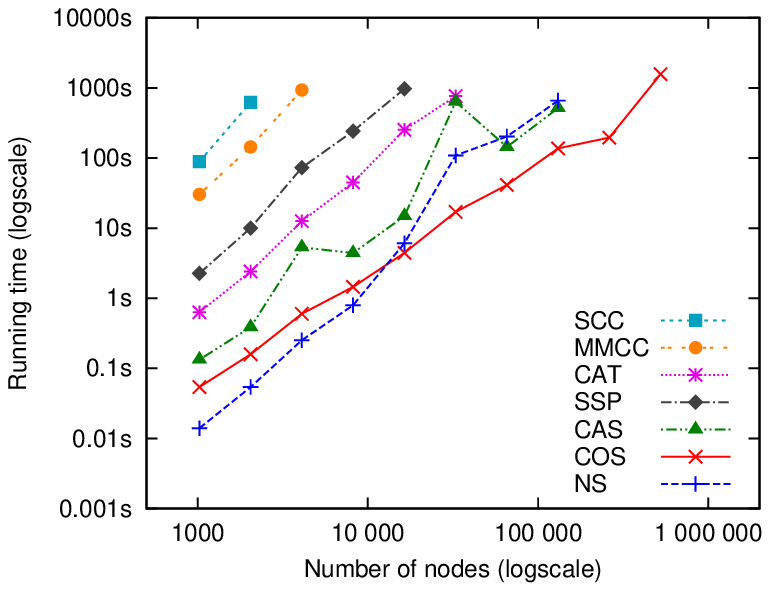}
  \end{minipage}
  \begin{minipage}{.5\textwidth}
    \centerline{\footnotesize\hspace{1em} GOTO-SR family ($m\approx n\sqrt{n}$)}\vspace{.5ex}
    \includegraphics[width=0.98\textwidth]{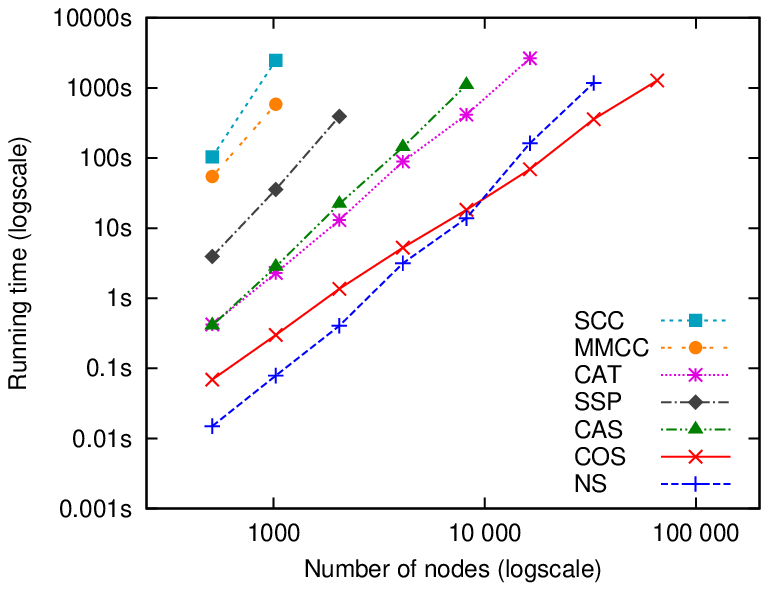}
  \end{minipage}
  \vspace{-1.5ex}
  \caption{Comparison of our implementations on GOTO instances}
  \label{fig:lemon_algs_goto}
\end{figure}

Figure~\ref{fig:lemon_algs_goto} and
Tables~\ref{tab:lemon_algs_goto_8} and \ref{tab:lemon_algs_goto_sr}
present the benchmark results for the GOTO families.
These problems indeed turned out to be much harder than the NETGEN instances
and the relative performance of the algorithms was also rather different than
in case of the NETGEN families.
Generally, COS and NS were the most efficient to solve these GOTO problems.
COS was clearly the best method for the largest instances, even for
the relatively dense GOTO\dash SR networks, which did not hold for
the NETGEN\dash SR instances.
Another notable difference compared to the NETGEN families
is that CAS was orders of magnitude faster than SSP.
It was quite efficient on GOTO\dash 8 instances, although its performance
was not stable.
Among the cycle-canceling algorithms, CAT was clearly the most efficient
similarly to the NETGEN problems.
However, MMCC was 3-4 times faster than SCC, which is in contrast
to the previous results.

\begin{table}[!htb]
  \centering
  \footnotesize
  \begin{tabular}{ccrrrrrrr}
   \hline \\[-2.25ex]
            $n$ & $deg$ &         SCC &        MMCC &         CAT &         SSP &         CAS &         COS &         NS \\
   \hline \\[-2.25ex]
    \pow{2}{10} &     8 &      88.85  &      30.39  &       0.63  &       2.27  &       0.13  &       0.05  &    \x{0.01} \\
    \pow{2}{12} &     8 &        $-$  &     931.54  &      12.62  &      72.93  &       5.34  &       0.60  &    \x{0.25} \\
    \pow{2}{14} &     8 &        $-$  &        $-$  &     253.54  &     971.03  &      14.87  &    \x{4.43} &       6.11  \\
    \pow{2}{16} &     8 &        $-$  &        $-$  &        $-$  &        $-$  &     143.92  &   \x{41.27} &     202.47  \\
    \pow{2}{18} &     8 &        $-$  &        $-$  &        $-$  &        $-$  &        $-$  &  \x{195.36} &        $-$  \\
    \pow{2}{20} &     8 &        $-$  &        $-$  &        $-$  &        $-$  &        $-$  &        $-$  &        $-$  \\
   \hline
  \end{tabular}
  \caption{Comparison of our implementations on GOTO-8 instances}
  \label{tab:lemon_algs_goto_8}
\end{table}

\begin{table}[!htb]
  \centering
  \footnotesize
  \begin{tabular}{crrrrrrrr}
   \hline \\[-2.25ex]
            $n$ & $deg$ &         SCC &        MMCC &         CAT &         SSP &         CAS &         COS &         NS \\
   \hline \\[-2.25ex]
    \pow{2}{10} &    32 &    2474.46  &     584.35  &       2.30  &      35.66  &       2.81  &       0.30  &    \x{0.08} \\
    \pow{2}{11} &    45 &        $-$  &        $-$  &      13.08  &     393.00  &      22.20  &       1.36  &    \x{0.41} \\
    \pow{2}{12} &    64 &        $-$  &        $-$  &      89.15  &        $-$  &     143.35  &       5.27  &    \x{3.16} \\
    \pow{2}{13} &    91 &        $-$  &        $-$  &     415.69  &        $-$  &    1101.38  &      18.23  &   \x{13.87} \\
    \pow{2}{14} &   128 &        $-$  &        $-$  &    2650.95  &        $-$  &        $-$  &   \x{69.37} &     163.37  \\
    \pow{2}{15} &   181 &        $-$  &        $-$  &        $-$  &        $-$  &        $-$  &  \x{358.44} &    1180.26  \\
    \pow{2}{16} &   256 &        $-$  &        $-$  &        $-$  &        $-$  &        $-$  & \x{1279.72} &        $-$  \\
   \hline
  \end{tabular}
  \caption{Comparison of our implementations on GOTO-SR instances}
  \label{tab:lemon_algs_goto_sr}
\end{table}

Table~\ref{tab:lemon_algs_road} presents the results obtained for the ROAD family.
As one would expect, the SSP algorithm was by far the fastest on these
special instances.
Our implementation of the CAS algorithm implies that it works exactly the same as
SSP on this family since all capacities are set to one.
Therefore, CAS is skipped in Table~\ref{tab:lemon_algs_road}.
The COS and NS algorithms performed an order of magnitude worse than SSP,
while the cycle-canceling algorithms were drastically slower.

\begin{table}[!htb]
  \centering
  \footnotesize
  \begin{tabular}{rrrrrrrrr}
   \hline \\[-2.25ex]
                $n$ & $deg$ &        SCC  &       MMCC  &        CAT  &        SSP  &        COS  &         NS  \\
   \hline \\[-2.25ex]
           9$\,$559 &  3.11 &       2.29  &     372.91  &       2.17  &    \x{0.01} &       0.14  &       0.04  \\
          49$\,$109 &  2.46 &      37.25  &        $-$  &      47.65  &    \x{0.10} &       1.47  &       0.69  \\
         116$\,$920 &  2.27 &     150.84  &        $-$  &     264.52  &    \x{0.26} &       4.61  &       2.97  \\
         261$\,$155 &  2.38 &    1984.30  &        $-$  &    1373.49  &    \x{0.96} &      15.23  &      14.14  \\
         519$\,$157 &  2.44 &        $-$  &        $-$  &        $-$  &    \x{3.29} &      35.87  &      41.38  \\
    1$\,$048$\,$506 &  2.53 &        $-$  &        $-$  &        $-$  &    \x{5.32} &      94.32  &     129.04  \\
    2$\,$073$\,$870 &  2.49 &        $-$  &        $-$  &        $-$  &   \x{21.99} &     238.57  &     744.36  \\
   \hline
  \end{tabular}
  \caption{Comparison of our implementations on ROAD instances}
  \label{tab:lemon_algs_road}
\end{table}

Finally, the performance results for the VISION family are presented
in Table~\ref{tab:lemon_algs_vision}.
We do not report running time for SCC and MMCC as they could not solve
these problems within the timeout limit of one hour.
COS performed clearly the best in these tests with the only exception of
the first instance.
Similarly to other families, the asymptotic behavior of NS was clearly
worse than that of COS.
CAS was superior to SSP, while CAT was even slower than SSP.

\begin{table}[!htb]
  \centering
  \footnotesize
  \begin{tabular}{rrrrrrr}
   \hline \\[-2.25ex]
                 $n$ & $deg$ &         CAT &         SSP &         CAS &         COS &         NS \\
   \hline \\[-2.25ex]
          245$\,$762 &  5.82 &     630.25  &     265.25  &     124.89  &      23.96  &   \x{21.06} \\
          491$\,$522 &  5.85 &    2304.01  &     970.49  &     622.22  &   \x{61.25} &     108.64  \\
          983$\,$042 &  5.88 &        $-$  &        $-$  &    1998.27  &  \x{186.42} &     531.59  \\
     1$\,$949$\,$698 &  5.91 &        $-$  &        $-$  &        $-$  &  \x{535.36} &    3420.76  \\
     3$\,$899$\,$394 &  5.92 &        $-$  &        $-$  &        $-$  & \x{1475.95} &        $-$  \\
   \hline
  \end{tabular}
  \caption{Comparison of our implementations on VISION instances}
  \label{tab:lemon_algs_vision}
\end{table}

Recall from Sections~\ref{sec:costscaling} and \ref{sec:networksimplex}
that the COS and NS algorithms have several variants.
These variants were also compared systematically to determine the default options.
Here we only present a representative selection of these results.

Tables~\ref{tab:lemon_cos_variants_netgen_8} and \ref{tab:lemon_cos_variants_goto_8}
compare the variants of the COS algorithm on the \mbox{NETGEN\dash 8} and
GOTO\dash 8 families, respectively.
COS\dash PR denotes the push-relabel variant,
COS\dash AR denotes the augment-relabel variant,
while COS denotes the partial augment-relabel variant, which is the default
implementation.
This latter technique was clearly faster than the
other two approaches on all kinds of problem instances.
The COS\dash AR variant performed similarly to COS\dash PR on some easy instances,
such as the NETGEN\dash 8 family,
but was an order of magnitude slower on some other instances, such as
the GOTO networks.
These results show that
COS\dash AR is not so robust than the other two methods, which is
in accordance with Goldberg's experiments in the maximum flow
context~\cite{Goldberg08PartialAug}.

\begin{table}[!htb]
  \centering
  \footnotesize
  \begin{tabular}{ccrrr}
   \hline \\[-2.25ex]
            $n$ & $deg$ &     COS-PR  &     COS-AR  &        COS  \\
   \hline \\[-2.25ex]
    \pow{2}{10} &     8 &       0.03  &    \x{0.02} &    \x{0.02} \\
    \pow{2}{12} &     8 &       0.17  &       0.14  &    \x{0.13} \\
    \pow{2}{14} &     8 &       0.94  &       1.11  &    \x{0.78} \\
    \pow{2}{16} &     8 &       6.37  &       5.72  &    \x{4.24} \\
    \pow{2}{18} &     8 &      35.17  &      28.00  &   \x{22.40} \\
    \pow{2}{20} &     8 &     176.87  &     179.13  &  \x{103.83} \\
    \pow{2}{22} &     8 &    1064.62  &     901.07  &  \x{615.42} \\
   \hline
  \end{tabular}
  \caption{Comparison of COS variants on NETGEN-8 instances}
  \label{tab:lemon_cos_variants_netgen_8}
\end{table}

\begin{table}[!htb]
  \centering
  \footnotesize
  \begin{tabular}{ccrrr}
   \hline \\[-2.25ex]
            $n$ & $deg$ &     COS-PR  &     COS-AR  &        COS  \\
   \hline \\[-2.25ex]
    \pow{2}{10} &     8 &       0.10  &       0.16  &    \x{0.05} \\
    \pow{2}{12} &     8 &       0.89  &       2.48  &    \x{0.60} \\
    \pow{2}{14} &     8 &       7.60  &      33.34  &    \x{4.43} \\
    \pow{2}{16} &     8 &      78.55  &     911.05  &   \x{41.27} \\
    \pow{2}{18} &     8 &     342.75  &        $-$  &  \x{195.36} \\
   \hline
  \end{tabular}
  \caption{Comparison of COS variants on GOTO-8 instances}
  \label{tab:lemon_cos_variants_goto_8}
\end{table}

We implemented five pivot rules for the NS method, which significantly
affect the efficiency of the algorithm.
Tables~\ref{tab:lemon_ns_variants_netgen_8} and \ref{tab:lemon_ns_variants_goto_8}
compare the overall performance of these strategies
on the NETGEN\dash 8 and GOTO\dash 8 families, respectively.
BE, FE, BS, CL, and AL denote the
best eligible, first eligible, block search, candidate list, and altering candidate list
pivot rules, respectively.
These results and many other experiments show that the BS and AL rules are
generally the most efficient.
On GOTO instances, the FE and CL rules also performed similarly to these methods,
but they were much slower in other cases, for example, on NETGEN instances.
Since the BS rule turned out to be slightly more robust than AL, it was selected to
be the default pivot strategy in our implementation.
The BE rule resulted in the worst performance although it yielded less iterations
than the other rules.

\begin{table}[!htb]
  \centering
  \footnotesize
  \begin{tabular}{ccrrrrr}
   \hline \\[-2.25ex]
            $n$ & $deg$ &      NS-BE  &      NS-FE  &      NS-BS  &      NS-CL  &      NS-AL  \\
   \hline \\[-2.25ex]
    \pow{2}{10} &     8 &       0.20  &    \x{0.01} &    \x{0.01} &    \x{0.01} &    \x{0.01} \\
    \pow{2}{12} &     8 &       3.40  &       0.14  &       0.05  &       0.06  &    \x{0.04} \\
    \pow{2}{14} &     8 &      60.47  &       3.64  &       0.54  &       1.02  &    \x{0.47} \\
    \pow{2}{16} &     8 &    1285.91  &     117.99  &       6.88  &      27.68  &    \x{6.41} \\
    \pow{2}{18} &     8 &        $-$  &        $-$  &     104.69  &     808.10  &   \x{98.97} \\
    \pow{2}{20} &     8 &        $-$  &        $-$  &  \x{799.26} &        $-$  &     800.36  \\
   \hline
  \end{tabular}
  \caption{Comparison of NS pivot rules on NETGEN-8 instances}
  \label{tab:lemon_ns_variants_netgen_8}
\end{table}

\begin{table}[!htb]
  \centering
  \footnotesize
  \begin{tabular}{ccrrrrr}
   \hline \\[-2.25ex]
            $n$ & $deg$ &      NS-BE  &      NS-FE  &      NS-BS  &      NS-CL  &      NS-AL  \\
   \hline \\[-2.25ex]
    \pow{2}{10} &     8 &       0.50  &    \x{0.01} &    \x{0.01} &    \x{0.01} &       0.02  \\
    \pow{2}{12} &     8 &       9.59  &       0.43  &    \x{0.25} &    \x{0.25} &       0.28  \\
    \pow{2}{14} &     8 &     151.92  &       6.84  &       6.11  &    \x{5.90} &       6.16  \\
    \pow{2}{16} &     8 &    3024.80  &     251.48  &  \x{202.47} &     216.21  &     220.16  \\
   \hline
  \end{tabular}
  \caption{Comparison of NS pivot rules on GOTO-8 instances}
  \label{tab:lemon_ns_variants_goto_8}
\end{table}

\subsection{Comparison to other solvers}
\label{sec:comparison_solvers}
The implementations presented in this paper
were also compared to the following widely known MCF solvers.
These codes were compiled using the same compiler and optimization level
as we used for our implementations.
The experiments were conducted using the default options of these solvers.

\begin{itemize}
  \item\textbf{CS2.}
  This is an authoritative implementation of the cost-scaling
  push-relabel algorithm.
  It was written by A.V.~Goldberg and B.~Cherkassky
  applying all improvements and heuristics described in
  \cite{Goldberg97EfficientImpl}.
  CS2 has been widely used as a benchmark for solving the MCF problem
  for a long time.
  We used the latest version, CS2~4.6, which is available
  from the IG~Systems, Inc.~\cite{CS2WEB}.

  \item\textbf{LEDA.}
  This is a comprehensive C++ library \cite{LEDA},
  which also provides an MCF solver in its \texttt{MIN\_COST\_FLOW()} procedure.
  This method implements the cost-scaling push-relabel algorithm, as well.
  We used version~5.0 of the LEDA library in our experiments.
  In fact, LEDA~5.1.1 was also tested, but it turned out to be slower than
  version~5.0.

  \item\textbf{MCFZIB.}
  This is the MCF code written by A. L\"obel \cite{Lobel96NetworkSimplex} at the
  Zuse Institute Berlin (ZIB).
  We denote this code as MCFZIB in order to differentiate it from the problem itself
  (similarly to the MCFClass project~\cite{MCFCLASS}).
  This solver features both a primal and a dual network simplex implementation,
  from which the former one is used by default as it is usually more efficient.
  We used the latest version 1.3, which is available at \cite{LobelMCF}.

  \item\textbf{RelaxIV.}
  This is a C++ translation of an authoritative implementation of the
  relaxation algorithm.
  The original FORTRAN code was written by D.P.~Bertsekas and P.~Tseng
  \cite{BertsekasTseng94RelaxIV} and is available at \cite{BertsekasNOC}.
  The C++ translation was made by A.~Frangioni and C.~Gentile
  at the University of Pisa and is available as part of the
  MCFClass project \cite{MCFCLASS}.
  This project provides a common C++ interface for several MCF solvers.
  Apart from RelaxIV, it also features CS2 and MCFZIB, but not their latest
  versions, thus we used these two solvers directly.
\end{itemize}

Our codes are part of an open source C++ optimization library,
\href{http://lemon.cs.elte.hu}{LEMON}, which is
available at \url{http://lemon.cs.elte.hu/}.

Tables~\ref{tab:comparison_netgen_8}, \ref{tab:comparison_netgen_sr},
\ref{tab:comparison_netgen_lo_8}, and \ref{tab:comparison_netgen_deg}
compare our implementations to the other four solvers
on the NETGEN instances.
As before, all codes were executed with a timeout limit of one hour
and the average running time over three different random instances is
reported for each problem size (in seconds).
Our COS code performed similarly to CS2 on NETGEN\dash 8 and
\mbox{NETGEN\dash SR} families, while it was slightly slower on the
other two NETGEN families.
The solver of the LEDA library was about two times slower than these
cost-scaling codes.
Furthermore, it failed to solve the largest instances due to a
number overflow error, which is denoted as ``\err'' in the tables.
Since LEDA has closed source, we could not eliminate this problem
by replacing the number types used by the algorithm with larger ones.
MCFZIB was typically slower than our NS code by a factor between
2 and 10, but they performed similarly on the
\mbox{NETGEN\dash LO\dash 8} instances.
RelaxIV was very efficient on these families.
It~was typically faster than all other codes for the largest instances,
while NS was the most efficient on the 
smaller networks and on the NETGEN\dash DEG family.

\begin{table}[!htb]
  \centering
  \footnotesize
  \begin{tabular}{cccrrcrrrr}
   \hline \\[-2.25ex]
                &       & & \multicolumn{2}{c}{LEMON} & & \multicolumn{4}{c}{Other implementations} \\
        \cline{4-5}\cline{7-10}\\[-2.25ex]
            $n$ & $deg$ & &        COS  &         NS  & &        CS2  &       LEDA  &     MCFZIB  &    RelaxIV  \\
   \hline \\[-2.25ex]
    \pow{2}{10} &     8 & &       0.02  &    \x{0.01} & &       0.02  &       0.03  &       0.02  &    \x{0.01} \\
    \pow{2}{12} &     8 & &       0.13  &    \x{0.05} & &       0.11  &       0.17  &       0.14  &       0.06  \\
    \pow{2}{14} &     8 & &       0.78  &    \x{0.54} & &       0.78  &       1.45  &       1.75  &    \x{0.54} \\
    \pow{2}{16} &     8 & &       4.24  &       6.88  & &       4.22  &       8.34  &      18.78  &    \x{3.58} \\
    \pow{2}{18} &     8 & &      22.40  &     104.69  & &      20.81  &       \err  &     207.29  &   \x{13.19} \\
    \pow{2}{20} &     8 & &     103.83  &     799.26  & &     103.25  &       \err  &    2985.08  &   \x{89.90} \\
    \pow{2}{22} &     8 & &     615.42  &        $-$  & &     566.32  &       \err  &        $-$  &  \x{419.38} \\
   \hline
  \end{tabular}
  \caption{Comparison to other solvers on NETGEN-8 instances}
  \label{tab:comparison_netgen_8}
\end{table}

\begin{table}[!htb]
  \centering
  \footnotesize
  \begin{tabular}{crcrrcrrrr}
   \hline \\[-2.25ex]
                &       & & \multicolumn{2}{c}{LEMON} & & \multicolumn{4}{c}{Other implementations} \\
        \cline{4-5}\cline{7-10}\\[-2.25ex]
            $n$ & $deg$ & &        COS  &         NS  & &        CS2  &       LEDA  &     MCFZIB  &    RelaxIV  \\
   \hline \\[-2.25ex]
    \pow{2}{10} &    32 & &       0.06  &    \x{0.01} & &       0.05  &       0.08  &       0.05  &       0.03  \\
    \pow{2}{11} &    45 & &       0.26  &    \x{0.05} & &       0.18  &       0.38  &       0.20  &       0.26  \\
    \pow{2}{12} &    64 & &       0.83  &    \x{0.21} & &       0.59  &       1.41  &       0.68  &       1.30  \\
    \pow{2}{13} &    91 & &       2.68  &    \x{0.73} & &       2.04  &       5.07  &       3.28  &       1.94  \\
    \pow{2}{14} &   128 & &       8.28  &    \x{3.55} & &       7.86  &      19.61  &      21.68  &       4.53  \\
    \pow{2}{15} &   181 & &      25.74  &      14.90  & &      29.00  &      72.48  &     111.74  &   \x{13.98} \\
    \pow{2}{16} &   256 & &     111.55  &      67.29  & &     104.25  &     274.06  &     634.38  &   \x{44.37} \\
   \hline
  \end{tabular}
  \caption{Comparison to other solvers on NETGEN-SR instances}
  \label{tab:comparison_netgen_sr}
\end{table}

\begin{table}[!htb]
  \centering
  \footnotesize
  \begin{tabular}{cccrrcrrrr}
   \hline \\[-2.25ex]
                &       & & \multicolumn{2}{c}{LEMON} & & \multicolumn{4}{c}{Other implementations} \\
        \cline{4-5}\cline{7-10}\\[-2.25ex]
            $n$ & $deg$ & &        COS  &         NS  & &        CS2  &       LEDA  &     MCFZIB  &    RelaxIV  \\
   \hline \\[-2.25ex]
    \pow{2}{10} &     8 & &       0.01  &    \x{0.00} & &       0.01  &       0.02  &       0.01  &       0.01  \\
    \pow{2}{12} &     8 & &       0.07  &    \x{0.02} & &       0.07  &       0.11  &       0.06  &       0.04  \\
    \pow{2}{14} &     8 & &       0.43  &    \x{0.19} & &       0.42  &       0.93  &       0.62  &       0.57  \\
    \pow{2}{16} &     8 & &       2.65  &    \x{2.11} & &       2.26  &       6.48  &       4.09  &       2.67  \\
    \pow{2}{18} &     8 & &      13.79  &      29.00  & &   \x{10.56} &       \err  &      28.31  &      18.73  \\
    \pow{2}{20} &     8 & &      68.41  &     293.49  & &   \x{54.48} &       \err  &     258.99  &      62.23  \\
    \pow{2}{22} &     8 & &     457.02  &    2482.50  & &     299.15  &       \err  &    2309.26  &  \x{229.67} \\
   \hline
  \end{tabular}
  \caption{Comparison to other solvers on NETGEN-LO-8 instances}
  \label{tab:comparison_netgen_lo_8}
\end{table}

\begin{table}[!htb]
  \centering
  \footnotesize
  \begin{tabular}{crcrrcrrrr}
   \hline \\[-2.25ex]
                &       & & \multicolumn{2}{c}{LEMON} & & \multicolumn{4}{c}{Other implementations} \\
        \cline{4-5}\cline{7-10}\\[-2.25ex]
            $n$ & $deg$ & &        COS  &         NS  & &        CS2  &       LEDA  &     MCFZIB  &    RelaxIV  \\
   \hline \\[-2.25ex]
    \pow{2}{12} &     2 & &       0.08  &    \x{0.02} & &       0.04  &       0.08  &       0.05  &       0.04  \\
    \pow{2}{12} &     8 & &       0.13  &    \x{0.05} & &       0.11  &       0.17  &       0.14  &       0.06  \\
    \pow{2}{12} &    32 & &       0.43  &    \x{0.13} & &       0.31  &       0.70  &       0.42  &       0.37  \\
    \pow{2}{12} &   128 & &       1.98  &    \x{0.39} & &       1.12  &       2.91  &       1.32  &       1.40  \\
    \pow{2}{12} &   512 & &       7.72  &    \x{1.31} & &       5.15  &      12.56  &       4.74  &       3.20  \\
    \pow{2}{12} &  2048 & &      47.08  &    \x{6.68} & &      32.72  &      69.55  &      18.52  &      12.81  \\
   \hline
  \end{tabular}
  \caption{Comparison to other solvers on NETGEN-DEG instances}
  \label{tab:comparison_netgen_deg}
\end{table}

\begin{figure}[!htb]
  \begin{minipage}{.5\textwidth}
    \centerline{\footnotesize\hspace{1em} GOTO-8 family ($m=8n$)}\vspace{.5ex}
    \includegraphics[width=0.98\textwidth]{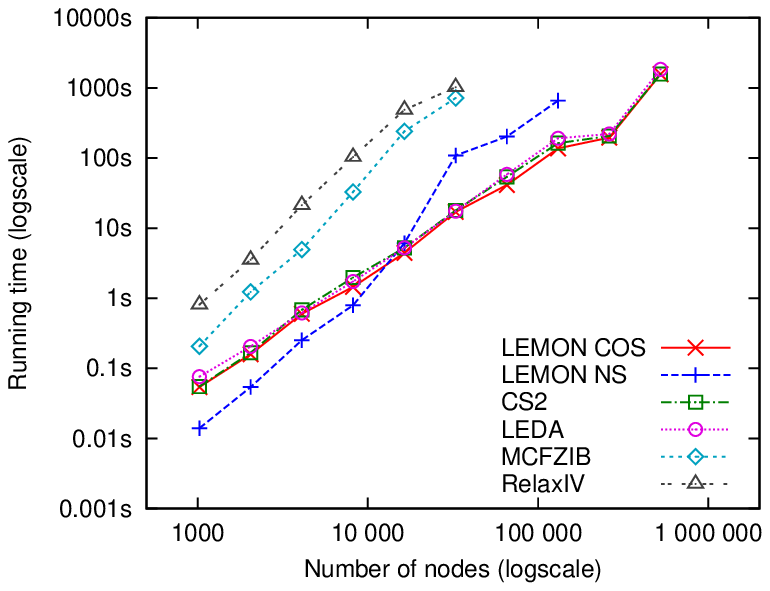}
  \end{minipage}
  \begin{minipage}{.5\textwidth}
    \centerline{\footnotesize\hspace{1em} GOTO-SR family ($m\approx n\sqrt{n}$)}\vspace{.5ex}
    \includegraphics[width=0.98\textwidth]{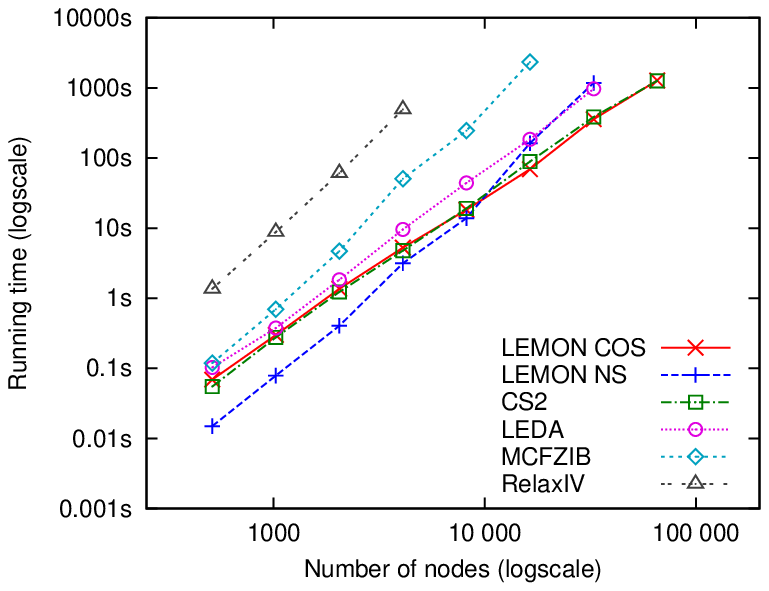}
  \end{minipage}
  \vspace{-1.5ex}
  \caption{Comparison to other solvers on GOTO instances}
  \label{fig:comparison_goto}
\end{figure}

The performance results for the GOTO families are presented in
Figure~\ref{fig:comparison_goto} and
Tables~\ref{tab:comparison_goto_8} and \ref{tab:comparison_goto_sr}.
In these tests, COS and CS2 also performed similarly and they were the
most efficient on the largest instances of both families.
LEDA was also similarly efficient on GOTO\dash 8 networks, but it was
2-3 times slower than CS2 and COS on the GOTO\dash SR family.
NS turned out to be orders of magnitude faster than the other network simplex
implementation, MCFZIB.
Similarly to the NETGEN families, NS was the most efficient algorithm on the
relatively small GOTO instances, but was substantially slower than
the cost-scaling codes on the large networks.
RelaxIV turned out to be very slow on these hard instances, which is in sharp
contrast to its efficiency on the NETGEN instances.

\begin{table}[!htb]
  \centering
  \footnotesize
  \begin{tabular}{cccrrcrrrr}
   \hline \\[-2.25ex]
                &       & & \multicolumn{2}{c}{LEMON} & & \multicolumn{4}{c}{Other implementations} \\
        \cline{4-5}\cline{7-10}\\[-2.25ex]
            $n$ & $deg$ & &        COS  &         NS  & &        CS2  &       LEDA  &     MCFZIB  &    RelaxIV  \\
   \hline \\[-2.25ex]
    \pow{2}{10} &     8 & &       0.05  &    \x{0.01} & &       0.06  &       0.08  &       0.21  &       0.81  \\
    \pow{2}{12} &     8 & &       0.60  &    \x{0.25} & &       0.69  &       0.62  &       4.94  &      21.29  \\
    \pow{2}{14} &     8 & &    \x{4.43} &       6.11  & &       5.23  &       5.24  &     239.67  &     487.23  \\
    \pow{2}{16} &     8 & &   \x{41.27} &     202.47  & &      54.05  &      58.50  &        $-$  &        $-$  \\
    \pow{2}{18} &     8 & &  \x{195.36} &        $-$  & &     206.48  &     221.14  &        $-$  &        $-$  \\
    \pow{2}{20} &     8 & &        $-$  &        $-$  & &        $-$  &        $-$  &        $-$  &        $-$  \\
   \hline                 
  \end{tabular}
  \caption{Comparison to other solvers on GOTO-8 instances}
  \label{tab:comparison_goto_8}
\end{table}

\begin{table}[!htb]
  \centering
  \footnotesize
  \begin{tabular}{crcrrcrrrr}
   \hline \\[-2.25ex]
                &       & & \multicolumn{2}{c}{LEMON} & & \multicolumn{4}{c}{Other implementations} \\
        \cline{4-5}\cline{7-10}\\[-2.25ex]
            $n$ & $deg$ & &        COS  &         NS  & &        CS2  &       LEDA  &     MCFZIB  &    RelaxIV  \\
   \hline \\[-2.25ex]
    \pow{2}{10} &    32 & &       0.30  &    \x{0.08} & &       0.28  &       0.38  &       0.70  &       8.81  \\
    \pow{2}{11} &    45 & &       1.36  &    \x{0.41} & &       1.23  &       1.83  &       4.70  &      60.89  \\
    \pow{2}{12} &    64 & &       5.27  &    \x{3.16} & &       4.78  &       9.57  &      50.65  &     492.38  \\
    \pow{2}{13} &    91 & &      18.23  &   \x{13.87} & &      19.16  &      44.24  &     246.38  &        $-$  \\
    \pow{2}{14} &   128 & &   \x{69.37} &     163.37  & &      89.32  &     184.97  &    2339.57  &        $-$  \\
    \pow{2}{15} &   181 & &  \x{358.44} &    1180.26  & &     385.14  &     973.49  &        $-$  &        $-$  \\
    \pow{2}{16} &   256 & &    1279.72  &        $-$  & & \x{1259.63} &        $-$  &        $-$  &        $-$  \\
   \hline
  \end{tabular}
  \caption{Comparison to other solvers on GOTO-SR instances}
  \label{tab:comparison_goto_sr}
\end{table}

Table~\ref{tab:comparison_road} presents the results for the ROAD family.
In this case, we also report the running time of our SSP implementation
since it was by far the most efficient method for this special family.
Apart from SSP, CS2 was typically the fastest to solve these instances.
COS was slower than CS2 by a factor of at most two, while
LEDA failed to solve the large instances due to overflow errors.
NS and MCFZIB were significantly slower than CS2 and COS on the large
networks. RelaxIV performed much worse than the other algorithms.

\begin{table}[!htb]
  \centering
  \footnotesize
  \begin{tabular}{rrc@{}rrrc@{\hspace{.5em}}rr@{\hspace{.75em}}r@{\hspace{.75em}}r}
   \hline \\[-2.25ex]
                    &       & & \multicolumn{3}{c}{LEMON} & & \multicolumn{4}{c}{Other implementations} \\
        \cline{4-6}\cline{8-11}\\[-2.25ex]
                $n$ & $deg$ & &        SSP  &        COS  &         NS  & &        CS2  &       LEDA  &     MCFZIB  &    RelaxIV  \\
   \hline \\[-2.25ex]                                     
           9$\,$559 &  3.11 & &    \x{0.01} &       0.14  &       0.04  & &       0.12  &       0.14  &       0.14  &       0.58  \\
          49$\,$109 &  2.46 & &    \x{0.10} &       1.47  &       0.69  & &       0.84  &       1.07  &       1.62  &      22.41  \\
         116$\,$920 &  2.27 & &    \x{0.26} &       4.61  &       2.97  & &       2.67  &       3.19  &       6.09  &     168.03  \\
         261$\,$155 &  2.38 & &    \x{0.96} &      15.23  &      14.14  & &       7.53  &       \err  &      27.25  &    2051.20  \\
         519$\,$157 &  2.44 & &    \x{3.29} &      35.87  &      41.38  & &      19.26  &       \err  &      81.07  &        $-$  \\
    1$\,$048$\,$506 &  2.53 & &    \x{5.32} &      94.32  &     129.04  & &      49.73  &       \err  &     197.54  &        $-$  \\
    2$\,$073$\,$870 &  2.49 & &   \x{21.99} &     238.57  &     744.36  & &     131.90  &       \err  &     992.32  &        $-$  \\
   \hline
  \end{tabular}
  \caption{Comparison to other solvers on ROAD instances}
  \label{tab:comparison_road}
\end{table}

The benchmark results for the VISION family are presented
in Table~\ref{tab:comparison_vision}.
In~these tests, CS2 was clearly the most efficient and COS was about
1.5 times slower than it.
All other algorithms performed significantly worse, including the third
cost-scaling code, LEDA.
NS was superior to MCFZIB and LEDA, but was less efficient
than CS2 and COS for the large networks.
Similarly to the GOTO and ROAD instances, RelaxIV was much slower than
the other solvers.

\begin{table}[!htb]
  \centering
  \footnotesize
  \begin{tabular}{rrcrrcrrrr}
   \hline \\[-2.25ex]
                   &       & & \multicolumn{2}{c}{LEMON} & & \multicolumn{4}{c}{Other implementations} \\
        \cline{4-5}\cline{7-10}\\[-2.25ex]
               $n$ & $deg$ & &        COS  &         NS  & &        CS2  &       LEDA  &     MCFZIB  &    RelaxIV  \\
   \hline \\[-2.25ex]
        245$\,$762 &  5.82 & &      23.96  &      21.06  & &   \x{19.78} &      54.62  &     131.66  &     904.74  \\
        491$\,$522 &  5.85 & &      61.25  &     108.64  & &   \x{44.18} &     207.51  &     333.20  &    3518.20  \\
        983$\,$042 &  5.88 & &     186.42  &     531.59  & &  \x{139.05} &    1250.56  &        $-$  &        $-$  \\
   1$\,$949$\,$698 &  5.91 & &     535.36  &    3420.76  & &  \x{348.28} &        $-$  &        $-$  &        $-$  \\
   3$\,$899$\,$394 &  5.92 & &    1475.95  &        $-$  & &  \x{916.47} &        $-$  &        $-$  &        $-$  \\
   \hline
  \end{tabular}
  \caption{Comparison to other solvers on VISION instances}
  \label{tab:comparison_vision}
\end{table}

Finally, on behalf of a brief overview of our experiments,
Table~\ref{tab:comparison_overview} presents running time results
for one representation of each problem family.
Those instances were selected that have about two million arcs.
The first part of the table contains running time in seconds,
while the second part reports normalized time results.
In the cases when the execution of an algorithm reached the timeout limit
of one hour, we report a lower bound on the ratio by which it would have been
slower than the fastest implementation.

\begin{table}[!htb]
  \centering
  \footnotesize
  \begin{tabular}{l@{}rc@{}rrc@{}rr@{\hspace{1em}}r@{\hspace{1em}}r}
   \textbf{Running time}\\
   \hline \\[-2.25ex]
                &       & & \multicolumn{2}{c}{LEMON} & & \multicolumn{4}{c}{Other implementations} \\
        \cline{4-5}\cline{7-10}\\[-2.25ex]
    Family      &             $m$ & &        COS  &         NS  & &        CS2  &       LEDA  &     MCFZIB  &    RelaxIV  \\
   \hline \\[-2.25ex]
    NETGEN-8    & 2$\,$097$\,$152 & &      22.40  &     104.69  & &      20.81  &       \err  &     207.29  &   \x{13.19} \\
    NETGEN-SR   & 2$\,$097$\,$152 & &       8.28  &    \x{3.55} & &       7.86  &      19.61  &      21.68  &       4.53  \\
    NETGEN-LO-8 & 2$\,$097$\,$152 & &      13.79  &      29.00  & &   \x{10.56} &       \err  &      28.31  &      18.73  \\
    NETGEN-DEG  & 2$\,$097$\,$152 & &       7.72  &    \x{1.31} & &       5.15  &      12.56  &       4.74  &       3.20  \\
    GOTO-8      & 2$\,$097$\,$152 & &  \x{195.36} &        $-$  & &     206.48  &     221.14  &        $-$  &        $-$  \\
    GOTO-SR     & 2$\,$097$\,$152 & &   \x{69.37} &     163.37  & &      89.32  &     184.97  &    2339.57  &        $-$  \\
    ROAD        & 2$\,$653$\,$624 & &      94.32  &     129.04  & &   \x{49.73} &       \err  &     197.54  &        $-$  \\
    VISION      & 2$\,$877$\,$382 & &      61.25  &     108.64  & &   \x{44.18} &     207.51  &     333.20  &    3518.20  \\
   \hline
   \\
   \textbf{Normalized time}\\
   \hline \\[-2.25ex]
                &       & & \multicolumn{2}{c}{LEMON} & & \multicolumn{4}{c}{Other implementations} \\
        \cline{4-5}\cline{7-10}\\[-2.25ex]
    Family      &             $m$ & &     COS  &      NS  & &     CS2  &    LEDA  &  MCFZIB  & RelaxIV  \\
   \hline \\[-2.25ex]
    NETGEN-8    & 2$\,$097$\,$152 & &    1.70  &    7.94  & &    1.58  &    \err  &   15.72  & \x{1.00} \\
    NETGEN-SR   & 2$\,$097$\,$152 & &    2.33  & \x{1.00} & &    2.21  &    5.52  &    6.11  &    1.28  \\
    NETGEN-LO-8 & 2$\,$097$\,$152 & &    1.31  &    2.75  & & \x{1.00} &    \err  &    2.68  &    1.77  \\
    NETGEN-DEG  & 2$\,$097$\,$152 & &    5.89  & \x{1.00} & &    3.93  &    9.59  &    3.62  &    2.44  \\
    GOTO-8      & 2$\,$097$\,$152 & & \x{1.00} &  $>$ 18  & &    1.06  &    1.13  &  $>$ 18  &  $>$ 18  \\
    GOTO-SR     & 2$\,$097$\,$152 & & \x{1.00} &    2.36  & &    1.29  &    2.67  &   33.73  &  $>$ 52  \\
    ROAD        & 2$\,$653$\,$624 & &    1.90  &    2.59  & & \x{1.00} &    \err  &    3.97  &  $>$ 38  \\
    VISION      & 2$\,$877$\,$382 & &    1.39  &    2.46  & & \x{1.00} &    4.70  &    7.54  &   79.63  \\
   \hline                               
  \end{tabular}                         
  \caption{Comparison of our implementations to other solvers on various problem instances with roughly
  		   the same number of arcs}
  \label{tab:comparison_overview}
\end{table}

Table~\ref{tab:comparison_overview} as well as the previous results
clearly demonstrate that CS2 and COS are the most robust
implementations and NS is the third one in terms of the overall performance.
MCFZIB is considerably slower than NS, but it is still robust.
LEDA performed well in several cases, but it often failed to solve
the large instances due to numerical problems.
RelaxIV is not robust at all as it turned out to be much slower than the other
implementations on all problem families except for the NETGEN networks.

\section{Conclusions}
\label{sec:conclusions}
In this paper, we have considered the minimum-cost flow (MCF) problem and
various solution methods along with experiments with their efficient
implementations.
The MCF problem plays a fundamental role in network flow theory and has a
wide range of applications.
Therefore, efficient implementations of MCF algorithms are essential in practice.

We implemented several algorithms for solving the MCF problem and thoroughly
experimented with many variants of them as well as with various
practical improvements and heuristics.
This work provides insight into these details and gives some guidelines
for implementing the considered algorithms efficiently.
An interesting novel result is the application of Goldberg's recent
partial augment-relabel idea \cite{Goldberg08PartialAug} in the cost-scaling
algorithm, which turned out to be a significant improvement.
Another widely used efficient algorithm is the network simplex method,
which was implemented using a quite efficient data structure and
various pivot strategies.
Moreover, three cycle-canceling algorithms and two augmenting path
algorithms were also implemented.

An extensive experimental evaluation was carried out to compare these
algorithms.
In general, the cost-scaling (COS) and the network simplex (NS) methods
turned out to be the most efficient and the most robust.
On relatively small instances (up to a few thousands of nodes),
NS was clearly the fastest algorithm.
However, COS significantly outperformed it on the
largest networks due to its better asymptotic behavior in terms of
the number of nodes.
We also remark that NS usually performed better than other methods
on rather dense networks, most likely because it is based on
maintaining a spanning tree data structure
and the tree update process depends only on the number of the nodes.
Apart from COS and NS, the other algorithms usually performed worse
and it turned out that their relative performance greatly depends on the
characteristics of the problem instance.
In certain cases, if the flow need not be
split into many paths, however, the augmenting path algorithms are
superior to other methods.

The presented implementations were systematically compared to
publicly available efficient MCF solvers, as well.
It turned out that our cost-scaling code is substantially more efficient and
more robust than that of the LEDA library~\cite{LEDA} and
it performs similarly to or slightly
slower than CS2 \cite{Goldberg97EfficientImpl, CS2WEB},
which is an authoritative implementation of this algorithm.
Our implementation of the network simplex method turned out to be
significantly faster than the other considered implementation
of this algorithm, the MCF solver \cite{Lobel96NetworkSimplex, LobelMCF}.
Furthermore, another well-known MCF solver, RelaxIV
\cite{BertsekasTseng94RelaxIV, BertsekasNOC, MCFCLASS} was also tested,
but it did not turn out to be robust at all.
It was orders of magnitude slower than the other codes on various
problem families, although it was rather efficient on particular
instances (namely, the NETGEN problems).

Our implementations are not standalone solvers, but they are part of a
versatile C++ network optimization library, \href{http://lemon.cs.elte.hu}{LEMON}
\cite{LEMON, DezsoJuttnerKovacs11Lemon} (\url{http://lemon.cs.elte.hu/}).
Therefore, these codes have the additional advantage that
they can easily be combined with various practical data structures and powerful
algorithms related to network optimization.
Furthermore, LEMON is an open source library that can be used in both
commercial and non-commercial software development under a
permissive license.
The authors believe that this library with its great variety of 
efficient algorithms is a viable alternative to the MCFClass project
\cite{MCFCLASS}, which features several publicly available MCF solvers
under a common C++ interface.

Finally, we remark some ideas for future work.
First, our implementation of the cost-scaling algorithm currently
does not incorporate the speculative arc fixing heuristic, which was suggested
by Goldberg \cite{Goldberg97EfficientImpl}.
We believe that the efficiency of this implementation could be further improved
by also applying this complicated technique.
Furthermore, a more comprehensive experimental study could be carried out
considering more problem families and more publicly available implementations
of MCF algorithms.

\vspace*{-6pt}
\section*{Acknowledgements}
The authors are grateful to A.V.~Goldberg and B.~Cherkassky,
D.P.~Bertsekas and P.~Tseng, and A. L\"obel for developing
CS2, RelaxIV, and MCF codes, respectively and for making these solvers
available for academic use.
We thank A. Frangioni and C. Gentile for developing the
MCFClass project including the C++ translation of the RelaxIV solver, which
was used in our experiments.
We also thank A. Frangioni for his help in using the MCFClass project.
Furthermore, we are grateful to A. J\"uttner and B. Dezs{\H o} 
for their useful suggestions related to the presented implementations.

Research was supported by grants (no. CNK~77780 and no. CK~80124) from the National Development
Agency of Hungary, based on a source from the Research and Technology Innovation Fund,
and by T\'AMOP grant 4.2.1/B-09/1/KMR-2010-0003.


\vspace*{-6pt}
\input{references}

\bigskip
\rightline{\emph{Received:  March 15, 2012 {\tiny \raisebox{2pt}{$\bullet$\!}} Revised: May 20, 2012}} 
\end{document}

%% file: references.tex

%% file: info41-5.bbl
\begin{thebibliography}{99}
{\small
\bibitem{AhujaEtAl92DoubleScaling}
R.K. \href{http://www.ise.ufl.edu/ahuja/}{Ahuja}, A.V. \href{http://www.avglab.com/andrew/}{Goldberg}, J.B. \href{http://jorlin.scripts.mit.edu/}{Orlin}, R.E. \href{http://www.cs.princeton.edu/~ret/}{Tarjan},
\newblock Finding minimum-cost flows by double scaling,
\newblock {\em Math. Program.} \textbf{53} (1992) 243--266.

\bibitem{AMO89NetworkFlows}
R.K. \href{http://www.ise.ufl.edu/ahuja/}{Ahuja}, T.L. \href{http://icampus.mit.edu/Symposium/Magnanti.aspx}{Magnanti}, J.B. \href{http://jorlin.scripts.mit.edu/}{Orlin},
\newblock Network flows,
\newblock in: G.L. Nemhauser, A.H.G.~Rinnooy Kan, M.J. Todd (eds), {\em
  Handbooks in Operations Research and Management Science}, Vol. 1, pp.  211--369. \href{www.elsevier.com/}{Elsevier}, 1989.

\bibitem{AMO93NetworkFlows}
R.K. \href{http://www.ise.ufl.edu/ahuja/}{Ahuja}, T.L. \href{http://icampus.mit.edu/Symposium/Magnanti.aspx}{Magnanti}, J.B. \href{http://jorlin.scripts.mit.edu/}{Orlin},
\newblock {\em Network Flows: Theory, Algorithms, and Applications},
\newblock \href{http://prenticehall.com/}{Prentice-Hall}, Inc., 1993.

\bibitem{BarahonaTardos89Note}
F.~Barahona, \'E. \href{http://www.cs.cornell.edu/People/eva/eva.html}{Tardos},
\newblock Note on {W}eintraub's minimum-cost circulation algorithm,
\newblock {\em SIAM J. Comput.} \textbf{18} (1989) 579--583.

\bibitem{BarrEtAl79TreeLabelling}
R.S. Barr, F.~\href{http://leeds-faculty.colorado.edu/glover/}{Glover}, D.~Klingman,
\newblock Enhancements to spanning tree labelling procedures for network
  optimization,  
\newblock {\em INFOR} \textbf{17} (1979) 16--34.

\bibitem{Bertsekas79Distributed}
D.P. \href{http://web.mit.edu/dimitrib/www/home.html}{Bertsekas},
\newblock A distributed algorithm for the assignment problem,
\newblock Working Paper, Laboratory for Information and Decision Systems, MIT,
  Cambridge, MA, 1979.

\bibitem{BertsekasNOC}
D.P. \href{http://web.mit.edu/dimitrib/www/home.html}{Bertsekas},
\newblock {N}etwork {O}ptimization {C}odes,
\newblock \url{http://web.mit.edu/dimitrib/www/noc.htm}, 1994.

\bibitem{BertsekasTseng94RelaxIV}
D.P. \href{http://web.mit.edu/dimitrib/www/home.html}{Bertsekas},  P.~\href{http://www.math.washington.edu/~tseng/personal.html}{Tseng},
\newblock {RELAX-IV}: {A} faster version of the relax code for solving minimum
  cost flow problems,
\newblock Technical Report LIDS-P-2276, Dept. of Electrical Engineering and
  Computer Science, MIT, Cambridge, MA, 1994.

\bibitem{BlandJensen92Computational}
R.G. \href{http://www.orie.cornell.edu/people/profile.cfm?netid=rgb6}{Bland}, D.L. Jensen.
\newblock On the computational behavior of a polynomial-time network flow
  algorithm,
\newblock {\em Math. Program.} \textbf{54} (1982) 1--39.

\bibitem{BradleyEtAl77Design}
G.~Bradley, G.~Brown, G.~Graves,
\newblock Design and implementation of large scale primal transshipment
  algorithms,
\newblock {\em Manag. Sci.} \textbf{24} (1977) 1--34.

\bibitem{BunnagelEtAl98EfficientImpl}
U.~B{\"u}nnagel, B.~Korte, J.~\href{http://www.or.uni-bonn.de/home/vygen/}{Vygen},
\newblock Efficient implementation of the {G}oldberg-{T}arjan minimum-cost flow
  algorithm,
\newblock {\em Opt. Methods and Software} \textbf{10} (1998) 157--174.

\bibitem{BusackerGowen60Procedure}
R.G. Busacker, P.J. Gowen,
\newblock A procedure for determining a family of minimum-cost network flow
  patterns,
\newblock Technical Report ORO-TP-15, Operations Research Office, The Johns
  Hopkins University, Bethesda, MD, 1960.

\bibitem{BusackerSaaty65Networks}
R.G. Busacker, T.L. Saaty,
\newblock {\em Finite Graphs and Networks: An Introduction with Applications},
\newblock McGraw-Hill, New York, NY, 1965.

\bibitem{COINOR}
{COIN-OR} -- {C}omputational {I}nfrastructure for {O}perations {R}esearch.
\newblock \url{http://www.coin-or.org/}, 2012.

\bibitem{CLRS09Algorithms}
T.H. \href{http://www.cs.dartmouth.edu/~thc/}{Cormen}, C.E. \href{http://people.csail.mit.edu/cel/}{Leiserson}, R.L. \href{http://people.csail.mit.edu/rivest/}{Rivest}, C.~\href{http://www.columbia.edu/~cs2035/}{Stein},
\newblock {\em Introduction to Algorithms},
\newblock The \href{http://mitpress.mit.edu/main/home/default.asp}{MIT} Press, 3rd edition, 2009.

\bibitem{CS2WEB}
{CS2} {S}oftware, {V}ersion {4.6}.
\newblock {IG} {S}ystems, {I}nc., \url{http://www.igsystems.com/cs2/}, 2009.

\bibitem{Cunningham79Theoretical}
W.H. Cunningham,
\newblock Theoretical properties of the network simplex method,
\newblock {\em Math. Oper. Res.} \textbf{4} (1979) 196--208.

\bibitem{Dantzig63LP}
G.B. Dantzig,
\newblock {\em Linear Programming and Extensions},
\newblock \href{http://press.princeton.edu/}{Princeton University Press}, Princeton, NJ, 1963.

\bibitem{Dasdan04Experimental}
A.~\href{http://dasdan.net/ali/}{Dasdan},
\newblock Experimental analysis of the fastest optimum cycle ratio and mean
  algorithms,
\newblock {\em ACM Trans. Des. Autom. Electron. Syst.} \textbf{9} (2004) 385--418.

\bibitem{DasdanGupta98FasterAlg}
A.~\href{http://dasdan.net/ali/}{Dasdan}, R.K. \href{http://mesl.ucsd.edu/gupta/}{Gupta},
\newblock Faster maximum and minimum mean cycle algorithms for system
  performance analysis,
\newblock {\em IEEE Trans. Computer-Aided Design} \textbf{17} (1998)  889--899.

\bibitem{DezsoJuttnerKovacs11Lemon}
B.~Dezs{\H o}, A.~\href{http://www.cs.elte.hu/~alpar/}{J\"uttner},  P.~\href{http://people.inf.elte.hu/kpeter/}{Kov\'acs},
\newblock {LEMON} -- an open source {C++} graph template library,
\newblock {\em Electron. Notes Theor. Comput. Sci.} \textbf{264} (2011) 23--45.

\bibitem{DIMACS1FTP}
\href{http://dimacs.rutgers.edu/}{DIMACS}, {N}etwork {F}lows and {M}atching: {F}irst {DIMACS} {I}mplementation
  {C}hallenge,
\newblock FTP site: \url{ftp://dimacs.rutgers.edu/pub/netflow/}, 2012.

\bibitem{DIMACS9WWW}
\href{http://dimacs.rutgers.edu/}{DIMACS}. {S}hortest {P}aths: {N}inth {DIMACS} {I}mplementation {C}hallenge,
\newblock WWW site: \url{http://www.dis.uniroma1.it/~challenge9/}, 2012.

\bibitem{EdmondsKarp72Theoretical}
J.~Edmonds, R.M. \href{http://www.eecs.berkeley.edu/~karp/}{Karp},
\newblock Theoretical improvements in algorithmic efficiency for network flow
  problems,
\newblock {\em J.~ACM} \textbf{19} (1972) 248--264.

\bibitem{EGRES}
{MTA-ELTE} {E}gerv\'ary {R}esearch {G}roup on {C}ombinatorial {O}ptimization
  ({EGRES}), 
\newblock \url{http://www.cs.elte.hu/egres/}, 2012.

\bibitem{FordFulkerson62Flows}
L.R. Ford,  D.R. Fulkerson,
\newblock {\em Flows in Networks},
\newblock \href{http://press.princeton.edu/}{Princeton University Press}, Princeton, NJ, 1962.

\bibitem{MCFCLASS}
A.~\href{http://www.di.unipi.it/~frangio/}{Frangioni}, C.~Gentile,
\newblock The {MCFClass} {P}roject,
\newblock \url{http://www.di.unipi.it/di/groups/optimize/Software/MCF.html},
  2011.

\bibitem{FrangioniManca06CostReopt}
A.~\href{http://www.di.unipi.it/~frangio/}{Frangioni}, A.~Manca, 
\newblock A computational study of cost reoptimization for min-cost flow
  problems, 
\newblock {\em INFORMS J. Comput.} \textbf{18} (2006) 61--70.

\bibitem{Frank11Connections}
A.~\href{http://www.cs.elte.hu/~frank/}{Frank},
\newblock {\em Connections in Combinatorial Optimization},
\newblock Oxford Lecture Series in Mathematics and Its Applications, Vol.~38.
  \href{http://www.oup.com/}{Oxford University Press}, 2011.

\bibitem{GabowTarjan89FasterScaling}
H.N. Gabow, R.E. \href{http://www.cs.princeton.edu/~ret/}{Tarjan},
\newblock Faster scaling algorithms for network problems,
\newblock {\em SIAM J. Comput.} \textbf{18}  (1989) 1013--1036.

\bibitem{GeorgiadisEtAl09Experimental}
L.~\href{http://www.icte.uowm.gr/lgeorg/}{Georgiadis}, A.V. \href{http://www.avglab.com/andrew/}{Goldberg}, R.E. \href{http://www.cs.princeton.edu/~ret/}{Tarjan}, R.F. \href{http://www.cs.princeton.edu/~rwerneck/}{Werneck},
\newblock An experimental study of minimum mean cycle algorithms,
\newblock In I.~Finocchi, J.~Hershberger, editors, {\em Proc. 6th
  International Workshop on Algorithm Engineering and Experiments}, ALENEX
  2009, pp 1--13, New York, NY, 2009. SIAM.

\bibitem{Goldberg95ScalingAlgs}
A.V. \href{http://www.avglab.com/andrew/}{Goldberg},
\newblock Scaling algorithms for the shortest paths problem,
\newblock {\em SIAM J. Comput.} \textbf{24} (1995) 494--504.

\bibitem{Goldberg97EfficientImpl}
A.V. \href{http://www.avglab.com/andrew/}{Goldberg},
\newblock An efficient implementation of a scaling minimum-cost flow algorithm,
\newblock {\em J.~Algorithms} \textbf{22}, 1  1--29.

\bibitem{Goldberg08PartialAug}
A.V. \href{http://www.avglab.com/andrew/}{Goldberg},
\newblock The partial augment-relabel algorithm for the maximum flow problem.
\newblock In {\em Proc. 16th Annual European Symposium on Algorithms}, ESA '08,
  pp. 466--477, Heidelberg, Germany, 2008. Springer-Verlag.

\bibitem{GoldbergKharitonov93Implementing}
A.V. \href{http://www.avglab.com/andrew/}{Goldberg}, M.~Kharitonov,
\newblock On implementing scaling push-relabel algorithms for the minimum-cost
  flow problem,
\newblock In D.S. Johnson, C.C. McGeoch (eds.) {\em {N}etwork {F}lows and
  {M}atching: {F}irst {DIMACS} {I}mplementation {C}hallenge}, {DIMACS} Series
  in Discrete Mathematics and Theoretical Computer Science, Vol.~12, pp.
  157--198. AMS, 1993.

\bibitem{GoldbergTarjan88NewApproach}
A.V. \href{http://www.avglab.com/andrew/}{Goldberg}, R.E. \href{http://www.cs.princeton.edu/~ret/}{Tarjan},
\newblock A new approach to the maximum-flow problem,
\newblock {\em J.~ACM} \textbf{35} (1988) 921--940.

\bibitem{GoldbergTarjan89CancelingCycles}
A.V. \href{http://www.avglab.com/andrew/}{Goldberg},  R.E. \href{http://www.cs.princeton.edu/~ret/}{Tarjan}.
\newblock Finding minimum-cost circulations by canceling negative cycles,
\newblock {\em J.~ACM} \textbf{36} (1989) 873--886.

\newpage\bibitem{GoldbergTarjan90SuccApprox}
A.V. \href{http://www.avglab.com/andrew/}{Goldberg}, R.E. \href{http://www.cs.princeton.edu/~ret/}{Tarjan}.
\newblock Finding minimum-cost circulations by successive approximation.
\newblock {\em Math. Oper. Res.} \textbf{15} (1990) 430--466.

\bibitem{GoldfarbHao92polynomial}
D.~\href{http://www.columbia.edu/~goldfarb/}{Goldfarb}, J.~Hao, 
\newblock Polynomial simplex algorithms for the minimum cost network flow
  problem, 
\newblock {\em Algorithmica} \textbf{8}  (1992) 145--160.

\bibitem{GoldfarbEtAl90Antistalling}
D.~\href{http://www.columbia.edu/~goldfarb/}{Goldfarb}, J.~Hao,  S.-R. Kai, 
\newblock Anti-stalling pivot rules for the network simplex algorithm, 
\newblock {\em Networks} \textbf{20} (1990) 79--91.

\bibitem{Grigoriadis86EfficientImpl}
M.D. Grigoriadis, 
\newblock An efficient implementation of the network simplex method, 
\newblock {\em Math. Program. Stud.} \textbf{26} (1986) 83--111.

\bibitem{HartmannOrlin93Finding}
M.~Hartmann, J.B. \href{http://jorlin.scripts.mit.edu/}{Orlin},
\newblock Finding minimum cost to time ratio cycles with small integral transit
  times, 
\newblock {\em Networks} \textbf{23} (1993) 567--574.

\bibitem{Howard60DynamicProg}
R.A. Howard,
\newblock {\em Dynamic Programming and Markov Processes}, 
\newblock The \href{http://mitpress.mit.edu/main/home/default.asp}{MIT} Press, 1960.

\bibitem{Iri60NewMethod}
M.~Iri, 
\newblock A new method for solving transportation-network problems, 
\newblock {\em J.~Oper. Res. Soc. Japan} \textbf{3} (1960) 27--87.

\bibitem{Jewell58Optimal}
W.S. Jewell,
\newblock Optimal flow through networks,
\newblock Technical Report Interim Technical Report No. 8, Operations Research
  Center, MIT, Cambridge, MA, 1958.

\bibitem{Karp78MinCycleMean}
R.M. \href{http://www.eecs.berkeley.edu/~karp/}{Karp},
\newblock A characterization of the minimum cycle mean in a digraph,
\newblock {\em Discrete Math.} \textbf{23} (1978) 309--311.

\bibitem{KellyOneill91NetworkSimplex}
D.J. Kelly, G.M. O'Neill,
\newblock The minimum cost flow problem and the network simplex solution
  method,
\newblock Master's thesis, University College, Dublin, 1991.

\bibitem{KenningtonHelgason80NetworkProg}
J.L. \href{http://www.smu.edu/Lyle/AboutUs/ContactsandDirectories/KenningtonJeffery}{Kennington},  R.V. \href{http://www.smu.edu/Lyle/AboutUs/ContactsandDirectories/HelgasonRichard}{Helgason},
\newblock {\em Algorithms for Network Programming},
\newblock John Wiley \& Sons, Inc., New York, NY, 1980.

\bibitem{KiralyKovacs10ExperimentalStudy}
Z.~\href{http://www.cs.elte.hu/~kiraly/}{Kir\'aly}, P.~\href{http://people.inf.elte.hu/kpeter/}{Kov\'acs},
\newblock An experimental study of minimum cost flow algorithms,
\newblock {\em Proc. 8th International Conference on Applied Informatics},
  Vol.~2., pp 227--235, Eger, Hungary, 2010.

\bibitem{Klein67PrimalMethod}
M.~Klein,
\newblock A primal method for minimal cost flows with applications to the
  assignment and transportation problems,
\newblock {\em Manag. Sci.} \textbf{14} (1967)  205--220.

\bibitem{KlingmanEtAl74Netgen}
D.~Klingman, A.~Napier, J.~Stutz,
\newblock {NETGEN}: {A} program for generating large scale capacitated
  assignment, transportation, and minimum cost flow network problems,
\newblock {\em Manag. Sci.} \textbf{20} (1974) 814--821.

\bibitem{KorteVygen12CombOpt}
B.~Korte \href{http://www.or.uni-bonn.de/index.eng.html}{}and J.~\href{http://www.or.uni-bonn.de/home/vygen/}{Vygen},
\newblock {\em Combinatorial Optimization: Theory and Algorithms},
\newblock Algorithms and Combinatorics, Vol.~21. Springer-Verlag, 5th edition,
  2012.

\bibitem{Kuhn55HungarianMethod}
H.W. Kuhn,
\newblock The {H}ungarian {M}ethod for the assignment problem,
\newblock {\em Naval Res. Logist. Quart.} \textbf{2} (1955)  83--97.

\bibitem{LEDA}
The {LEDA} library, {V}ersion 5.0.
\newblock {A}lgorithmic {S}olutions {S}oftware {GmbH},
  \url{http://www.algorithmic-solutions.com/leda/}, 2004.

\bibitem{LEMON}
{LEMON} -- {L}ibrary for {E}fficient {M}odeling and {O}ptimization in
  {N}etworks,
\newblock \url{http://lemon.cs.elte.hu/}, 2012.

\bibitem{LEMONDOC}
{LEMON} {D}ocumentation, {V}ersion {1.2.3},
\newblock \url{http://lemon.cs.elte.hu/pub/doc/1.2.3/}, 2012.

\newpage\bibitem{Lobel96NetworkSimplex}
A.~\href{http://www.zib.de/loebel/}{L\"obel},
\newblock Solving large-scale real-world minimum-cost flow problems by a
  network simplex method,
\newblock Technical Report SC 96-7, Zuse Institute Berlin (ZIB), Berlin,
  Germany, 1996.

\bibitem{LobelMCF}
A.~\href{http://www.zib.de/loebel/}{L\"obel},
\newblock {MCF} {V}ersion 1.3 -- {A} network simplex implementation.
\newblock   \href{http://typo.zib.de/opt-long_projects/Software/Mcf/}{www.zib.de}, 2004.

\bibitem{Mulvey78Pivot}
J.M. Mulvey.
\newblock Pivot strategies for primal-simplex network codes.
\newblock {\em J.~ACM} \textbf{25} (1978) 266--270.

\bibitem{Orlin93Faster}
J.B. \href{http://jorlin.scripts.mit.edu/}{Orlin},
\newblock A faster strongly polynomial minimum cost flow algorithm,
\newblock {\em Oper. Res.} \textbf{41} (1993) 338--350.

\bibitem{Orlin97PolyPrimal}
J.B. \href{http://jorlin.scripts.mit.edu/}{Orlin},
\newblock A polynomial time primal network simplex algorithm for minimum cost
  flows,
\newblock {\em Math. Program.}  \textbf{78} (1997) 109--129.

\bibitem{OrlinEtAl93PolyDual}
J.B. \href{http://jorlin.scripts.mit.edu/}{Orlin}, S.A. \href{http://troll-w.stanford.edu/plotkin/}{Plotkin},  \'E. \href{http://www.cs.cornell.edu/People/eva/eva.html}{\href{http://www.cs.cornell.edu/People/eva/eva.html}{Tardos}},
\newblock Polynomial dual network simplex algorithms,
\newblock {\em Math. Program.} \textbf{60} (1993) 255--276 1993.

\bibitem{Rock80Scaling}
H.~R\"ock.
\newblock Scaling techniques for minimal cost network flows,
\newblock In U.~Pape, editor, {\em Discrete Structures and Algorithms}, pp.
  181--191, M\"unchen, 1980. Carl Hanser.

\bibitem{Schrijver03CombOpt}
A.~\href{http://homepages.cwi.nl/~lex/}{Schrijver},
\newblock {\em Combinatorial Optimization: Polyhedra and Efficiency},
\newblock Algorithms and Combinatorics, Vol.~24. Springer-Verlag, 2003.

\bibitem{SleatorTarjan83DynamicTree}
D.D. \href{http://www.cs.cmu.edu/~sleator/}{Sleator}, R.E. \href{http://www.cs.princeton.edu/~ret/}{Tarjan},
\newblock A data structure for dynamic trees,
\newblock {\em J.~Comput. System Sci.} \textbf{26} (1983) 362--391.

\bibitem{SokkalingamEtAl00NewPoly}
P.T. Sokkalingam, R.K. \href{http://www.ise.ufl.edu/ahuja/}{Ahuja},  J.B. \href{http://jorlin.scripts.mit.edu/}{Orlin},
\newblock New polynomial-time cycle-canceling algorithms for minimum-cost
  flows,
\newblock {\em Networks} \textbf{36} (2000) 53--63.

\bibitem{Tardos85StronglyPoly}
\'E. \href{http://www.cs.cornell.edu/People/eva/eva.html}{Tardos},
\newblock A strongly polynomial minimum cost circulation algorithm,
\newblock {\em Combinatorica} \textbf{5} (1985) 247--255.

\bibitem{Tarjan91Efficiency}
R.E. \href{http://www.cs.princeton.edu/~ret/}{Tarjan},
\newblock Efficiency of the primal network simplex algorithm for the
  minimum-cost circulation problem, 
\newblock {\em Math. Oper. Res.} \textbf{16} (1991)  272--291.

\bibitem{Tarjan97DynamicTreesNetwork}
R.E. \href{http://www.cs.princeton.edu/~ret/}{Tarjan},
\newblock Dynamic trees as search trees via euler tours, applied to the network
  simplex algorithm,
\newblock {\em Math. Program.} \textbf{78} (1997) 169--177.

\bibitem{Tomizawa71Techniques}
N.~Tomizawa, 
\newblock On some techniques useful for solution of transportation network
  problems, 
\newblock {\em Networks} \textbf{1} (1971) 173--194.

}\end{thebibliography}
